\documentclass[nohyper,11pt,letterpaper]{JHEP3}
\usepackage[dvips]{epsfig}
\usepackage{amsmath}

\newcommand{\be}{\begin{equation}}
\newcommand{\ee}{\end{equation}}
\newcommand{\ben}{\begin{eqnarray}}
\newcommand{\een}{\end{eqnarray}}
\newcommand{\ba}{\begin{eqnarray}}
\newcommand{\ea}{\end{eqnarray}}
\newcommand{\nn}{\nonumber \\}

\newcommand{\beq}{\begin{equation}}
\newcommand{\eeq}{\end{equation}}

\newcommand{\mt}[1]{\textrm{\tiny #1}}

\newcommand{\pf}{partition function \ }

\newcommand{\un}{{$U(N)$}}
\newcommand{\ra}{\rightarrow}
\newcommand{\seff}{S_\mt{eff}}
\newcommand{\p}{\partial}
\newcommand{\hm}{q}
\newcommand{\hma}{p}
\newcommand{\vr}{\xi}

\newcommand{\tx}{\tilde {\xi}}
\newcommand{\bi}{\begin{itemize}}
\newcommand{\ei}{\end{itemize}}
\newcommand{\ii}{\item}

\newcommand{\yt}{Young Tableaux}

\newcommand{\me}{\mt{\large e}}

\newcommand{\ml}{\ln}
\newcommand{\mT}{\mt{\large T}}

\newcommand{\mtr}{\mt{\large Tr}}

\newcommand{\lb}{\left(}
\newcommand{\rb}{\right)}
\newcommand{\ltb}{\left [}
\newcommand{\rtb}{\right ]}
\newcommand{\bc}{\begin{center}}
\newcommand{\ec}{\end{center}}
\newcommand{\mU}{\mit\Upsilon}

\title{Free Fermions and Thermal AdS/CFT}

\author{Suvankar Dutta$^a$ and Rajesh Gopakumar$^b$ \\
$^{a,b}$Harish-Chandra Research Institute, Chhatnag Road, Jhusi,
  Allahabad 211 019, INDIA\\
E-mail:\ $^a$suvankar@mri.ernet.in, \ $^b$gopakumr@mri.ernet.in}

\abstract{The dynamics of finite temperature $U(N)$ gauge theories on
$S^3$ can be described, at weak coupling, by an effective unitary
matrix model. Here we present an exact solution to these models, for
any value of $N$, in terms of a sum over representations. Taking the
large $N$ limit of this solution provides a new perspective on the
deconfinement transition which is supposed to be dual to the
Hawking-Page transition. The large $N$ phase transition manifests
itself here in a manner similar to the Douglas-Kazakov phase
transition in $2d$ Yang-Mills theory. We carry out a complete analysis
of the saddle representation in the simplest case involving only the
order parameter ${\rm Tr}U$. We find that the saddle points
corresponding to thermal $AdS$, the small black hole and the large
black hole can all be described in terms of free fermions. They all
admit a simple phase space description {\it a la} the BPS geometries
of Lin, Lunin and Maldacena.}


\preprint{}

\begin{document}{\vskip 1cm}

\section{Introduction} \label{intro}

Though we know of many instances where a gauge theory is a holographic
description of a gravitational theory, we are yet to understand the
precise way in which a local diffeomorphism invariant theory in one
higher dimension is encoded in the dynamics of the gauge theory. In a
sense, the redundancy of diffeomorphisms has been largely eliminated
in the gauge theory description. But this has come at the cost of
losing information about the locality of the bulk description. Is
there a natural way in the gauge theory to restore the redundancies
which characterise the geometrical description of the bulk?

A partial hint comes from the beautiful work of Lin, Lunin and
Maldacena \cite{llm} who showed that the geometry of a class of
half-BPS solutions of the bulk theory is completely fixed by
specifying a single function of two of the bulk coordinates. This
function, which takes values either zero or one in the entire two
dimensional plane, was identified with the phase space distribution of
free fermions describing the half BPS dynamics in the gauge theory. In
other words, the {\it configuration space} of the bulk, with its
redundancies, was identified with the {\it phase space} of the
boundary degrees of freedom. In fact, the quantisation of the space of
BPS configurations on the gravity side agrees with those of the free
fermions \cite{mandal}\cite{grant}.  Notice that the fermionic phase
space description is also a redundant one since it is the shape of the
perimeter of the "filled fermi" sea that completely determines
everything. The phase space picture therefore appears to be a step in
the right direction.

However, the half-BPS case seems to be very special and the picture of
free fermions is not likely to be generally applicable. It is
therefore a bit of a surprise that, in this paper, we find a similar
free fermion phase space description in the non-supersymmetric context
of finite temperature AdS/CFT. It is very well known \cite{witten}
that the thermal partition function of the gauge theory exhibits a
behaviour which is qualitatively similar to the Hawking-Page \cite{hp}
phase diagram on the gravity side.  In particular, we find a free
fermionic description in the weakly coupled gauge theory, for each of
the saddle points that correspond to Thermal AdS, the (unstable)
"small" AdS Schwarschild black hole as well as the "big" AdS black
hole.  In each case there is a simple region in phase space which is
the filled fermi sea.

Our starting point is the effective unitary matrix model that
describes the holonomies of the Polyakov loop at weak gauge coupling
\cite{sundborg}\cite{shiraz}.  As was argued in \cite{shiraz}, in the
free $U(N)$ Yang-Mills theory on $S^3$ at finite temperature, all
modes are massive and can be exactly integrated out, except for the
zero mode of $A_0$. The dynamics of this interacting mode is naturally
expressed in terms of a unitary matrix model for the holonomy $U$,
along the thermal $S^1$.  At weak coupling, we can continue to
integrate out all the other modes and end up with (a more complicated)
effective matrix model for $U$.  These matrix models have been well
analysed, in the large $N$ limit, in terms of the collective field
$\sigma(\theta)$ which is the eigenvalue density of $U$. They have
been shown to exhibit a phase structure which describes the
deconfinement transition and is qualitatively very similar to that of
the Hawking-Page description of AdS gravity at finite temperature
\cite{shiraz}\cite{hong}\cite{spenta}.

In this paper, we present an exact solution to the partition function
of these matrix models, which is valid for any finite $N$. The answer
is in terms of characters of the conjugacy classes of the symmetric
group with a sum over different representations and classes. While
explicit, the expressions are, in general, quite complicated.  At
large $N$ we expect the answer to show the non-analytic behaviour, as
one varies the temperature, which is characteristic of a phase
transition. This is seen in our expressions from the fact that at
large $N$, there is a dominant saddle point in the sum over
representations. The nature of this saddle point exhibits non-analytic
jumps as one varies the temperature. This is similar to how the large
$N$ phase transition of Douglas-Kazakov \cite{dk}, in $2d$ Yang-Mills
theory, manifests itself.

The quantitative method of analysis, as in $2d$ Yang-Mills, introduces
a density $u(h)$ for the Young tableaux that label the
representations. This essentially measures the number of boxes in the
rows of the tableaux. One can write an effective action for $u(h)$ in
the large $N$ limit and study its saddle points. We do this analysis
very explicitly for the simplest and physically important case\footnote{The 
$(a,b)$ model studied in \cite{spenta}
falls, for instance, in this class.} where
one has only terms involving $\mtr U$ and $\mtr
U^{\dagger}$. 
The actual saddle point equations
are close to that of models studied by Kazakov, Staudacher and Wynter
\cite{ksw}, though those cases did not exhibit a phase transition.  In
our case, one finds, not surprisingly, exactly the phase diagram
obtained by the usual eigenvalue density analysis.

However, what is of interest in the present analysis, is the nature of
the saddle point representations $u_0(h)$, in both the low and high
temperature phases. It turns out that they bear a simple relation to
the saddle point eigenvalue densities $\sigma_0(\theta)$. Essentially
the two turn out to be functional inverses of each other. The best
way, in fact, to state the relation between the two is to view
$(h,\theta)$ as coordinates on a two dimensional phase space and
define an appropriate region $R$ with {\it constant} fermion density
$\rho(h,\theta)={1\over 2\pi}$ in its interior and zero outside. It
then turns out that 
\ben \label{phsp} 
\int \rho(h,\theta)dh &=& \sigma_0(\theta) \nn 
\int \rho(h,\theta) d\theta &=& u_0(h).  
\een 
In other words, the region $R$ is determined by the shape
$h_0(\theta)$ which is obtained from inverting the equation
$\theta=\pi u_0(h)$.

It turns out that the region $R$ corresponding to thermal $AdS$ is
given by the unit disk in phase space. This is indeed what one also
obtains in the LLM picture \cite{llm} for the global $AdS$ spacetime.
The regions corresponding to the small black hole and and the big
black hole are more complicated kidney-shaped geometries as shown in
Figs.7-9. The shape of these regions is not modified in functional form
when one includes perturbative corrections in terms of an effective
action involving only the relevant winding number one modes. Thus at
least in the weak coupling expansion this geometry of the phase space
distribution is robust and therefore can be expected to capture some
essential features of the corresponding bulk geometries. It would be
very interesting to learn what these features might be. In particular,
it is natural to ask whether there is a direct translation into a
supergravity solution like in the LLM case. Gaining an understanding
of these points might help us learn why the matrix models capture the
dynamics of the gravity phase transition so well.

The plan of the paper is as follows. In the next section
(Sec.\ref{secrev}) we review the unitary matrix models that describe
the finite temperature dynamics at weak coupling. We also recapitulate
the results that follow from a large $N$ analysis in terms of the
eigenvalue density $\sigma(\theta)$ and the correspondence with the
phase diagram on the gravity side. In Sec.\ref{secexatsol} we write
down the exact finite $N$ solution to the models at zero coupling and
also show how the method of solution generalises to the weakly coupled
case. In Sec.\ref{seclargeN} we analyse the large $N$ limit of the
exact solution for models involving only $\mtr U$ and $\mtr
U^{\dagger}$. We do this in terms of the Young Tableaux density $u(h)$
and find a phase transition as expected. We find the expressions for
the saddle points $u_0(h)$ that dominate at both low and high
temperature and compute their free energies to find agreement with the
results of the eigenvalue density analysis.  In Sec.\ref{secevyteqv},
we show how the results of Sec.\ref{seclargeN} imply a relation of
$u_0(h)$ with the saddle point eigenvalue densities
$\sigma_0(\theta)$. We show how this relation can be simply understood
in terms of a free fermionic phase space picture in
Sec.\ref{secfrefer}.  In Sec.\ref{con} we close with various comments
on the possible implications of these results which need to be fleshed
out in the future. Appendices contain details of some of the
calculations as well as some generalisations.

\section{The Finite Temperature Partition Function for Gauge Theories  
on $S^3$} \label{secrev}

\subsection{The Effective Action for the Holonomy} \label{effacnholo}

In the AdS/CFT correspondence the four dimensional gauge theory lives
on the boundary $S^3$ (together with the $R$ direction for time) of
the $AdS_5$ spacetime. In the finite temperature version, the
(Euclidean) gauge theory now has a thermal $S^1$ instead of
$R$. Studying the dynamics of the thermal gauge theory on $S^3 \times
S^1$ offers some important simplifications. In the free gauge theory
(defined as the $\lambda=g_{YM}^2N \rightarrow 0$ limit), most of the
modes are massive with a scale set by the radius of the $S^3$. There
is a single massless mode which is the zero mode of the temporal
component of the gauge field.  
\be\label{azero} \alpha={1\over
V_{S^3}}\int_{S^3} A_0.  
\ee 
This mode is therefore strongly
self-interacting even at arbitrarily weak 'tHooft coupling $\lambda$.
Consequently, one can, in the free theory, exactly integrate out all the
massive modes and obtain an {\it exact} effective action for the mode
$\alpha$.  This analysis was carried out in \cite{shiraz} and one
obtains a unitary matrix model in terms of the holonomy\footnote{This
unitary matrix model representation of the finite temperature
partition function was given first by \cite{sundborg} based on
enumeration of states in the free theory. See also \cite{hallin}.}
\be\label{hol} U=e^{i\beta \alpha}, 
\ee 
where $\beta={1\over T}$ is
the radius of the thermal circle\footnote{The unitary matrix model for
the free gauge theory was obtained earlier by Sundborg \cite{sundborg} by
counting states of the free theory. See also \cite{hallin}.}.
 
One finds that the gauge theory \pf(with $U(N)$ gauge group and
restricting to adjoint matter fields) on $S^3 \times S^1$ is given by,
\be \label{freepf} Z(\beta) = \int [dU] \exp\left[ \sum_{n=1}^{\infty}
{a_n(T) \over n} \mtr(U^n) \mtr(U^{\dagger n}) \right], 
\ee 
where the coefficients $a_n(T)$ are given, in terms of
$x=e^{-{\beta}}$, by 
\be a_n(T) = z_\mt{B} (x^n) + (-1)^{n+1}
z_\mt{F}(x^n).  
\ee 
Here $z_\mt{B}(x)$ and $z_\mt{F}(x)$ are single
particle partition functions of the bosonic and fermionic.modes
respectively. They completely capture the field content of the gauge
theory.  The explicit expressions for $z_\mt{B}(x)$ and $z_\mt{F}(x)$,
for fields of different spin are given in \cite{shiraz}.

The above expression was derived at zero coupling where one has only a
one loop contribution from integrating out all the massive modes. For
weak 't Hooft coupling, one may continue to integrate out the massive
modes and obtain a more general (and more complicated effective action
for the holonomy $U$). The structure of the effective action is now
\cite{shiraz}
\be\label{weakpf}
Z(\beta, \lambda) = \int
[dU] \exp S_{eff}(U)
\ee
where
\be\label{weakacn}
S_{eff}(U)=\sum_{\{ n_i\} }a_{\{ n_i\} }
(\lambda, T, N){1\over N^k}\prod_{i=1}^k\mtr U^{n_i}
\ee
with the integers $n_i$ obeying $\sum_i n_i=0$ and the coefficients
$a_{\{ n_i\} }$, of a term with $k$ traces, making their first
appearance at $(k-1)$ loops in perturbation theory and consequently
having a planar contribution starting with $\lambda^{k-2}$.

Therefore, in perturbation theory, all the non-trivial low energy
dynamics of the
finite temperature theory on $S^3$ is captured by this unitary matrix
model. It is the properties of this general class of unitary matrix
models that we will study in this paper.

However, one can make a further important simplification.  The order
parameter for the large $N$ phase transition exhibited by these models
(reviewed in the next subsection) is ${\rm Tr}U$. Consequently, one
can also imagine integrating out all the ${\rm Tr}U^n$ (with $n \neq
\pm 1$) and obtaining an effective action purely in terms of ${\rm
Tr}U{\rm Tr}U^{\dagger}$.  This is not easy to carry out
explicitly. Therefore one can consider toy models of the form
\cite{spenta}
\be\label{genZ}
Z = \int [dU] \me^{N^2 \seff(x)} \ , \quad x={1 \over N^2} \mtr U
\mtr U^{\dagger} \ ,
\ee
where 
\be\label{sefftrunc}
\seff(U)= a_1(\lambda, \mT) \mtr U \mtr
U^{\dagger} + {b_1(\lambda, \mT) \over N^2} (\mtr U  \mtr
U^{\dagger})^2 
+ {c_1(\lambda, \mT) \over N^4} \ (\mtr U \mtr
U^{\dagger})^3 + \ \cdot \cdot \cdot 
\ee
with $S(x)$ being convex and $S'(x)$ being concave. 
The simplest such model is the so-called $(a,b)$ model \cite{spenta}
in which one keeps only the 
first two coefficients in the $\seff$ given in Eq.\ref{sefftrunc}. 
\be\label{Zab}
Z(a_1,b_1) = \int [dU] \exp \ltb a_1\mtr  U \mtr U^{\dagger} + {b_1
\over N^2} \lb \mtr U \mtr U^{\dagger} \rb^2 \rtb ,
\ee
where $a_1$ and $b_1$ are functions of temperature $T$ and $\lambda$.

\subsection{Eigenvalue Density Analysis at Large $N$} \label{secevana}

The above unitary matrix models can be analysed using standard
techniques in the large $N$ limit. We briefly review the results
\cite{shiraz}\cite{spenta} in this subsection.

One introduces the eigenvalue density 
\be
\sigma(\theta)={1\over N}\sum_{i=1}^N\delta(\theta-\theta_i)
\ee
where the holonomy matrix $U$ takes the diagonal form 
\be
U=diag(e^{i\theta_i}).
\ee

Let us start with the free \pf Eq.\ref{freepf}. It can be expressed in
terms of a functional $S[\sigma(\theta)]$
\be
Z(\beta)=\int [D\sigma] \me^{N^2 S[\sigma(\theta)]},
\ee
where 
\be\label{freefunc} 
S[\sigma(\theta)]= \int d\theta_1 \int d\theta_2 \sigma(\theta_1)
\sigma(\theta_2) V(\theta_1 -\theta_2). 
\ee
Here, the two body potential $V(\theta, T)$ is given by
\be 
V(\theta, T) = \ln[2] + \sum_{n=1}^{\infty} {1\over n} 
(1 - a_n(T)) \cos(n\theta). 
\ee 

The general case for arbitrary $a_n$ is actually quite cumbersome to
analyse. A self consistent method was given in \cite{shiraz}. See also
\cite{spenta2} for a more general method.  As mentioned above the
crucial order parameter is ${\rm Tr} U$, therefore we will often
concentrate on the case where we keep only the terms with $\mtr U,
\mtr U^{\dagger}$. In other words, we set $a_n=0$ for $n>1$.  For this
case\footnote{One can estimate that $a_n \sim 0$ for $n>1$, at the
phase transition temperature.  Even at higher temperatures, the
contributions of the higher $a_n$ is typically small.}  where only
$a_1\neq 0$, one can explicitly obtain the saddle points for the above
functional \ref{freefunc}. One finds the following

\bi
\item
For $a_1<1$, the minimum action is for the eigenvalue density
\be\label{sigunfrm}
\sigma(\theta)={1\over 2\pi}.
\ee
The free energy is zero for this configuration (to order $N^2$). 
\item
For $a_1=1$, there is a continuous family of minimum action
configurations (labeled by a parameter $\vr$) for which the
eigenvalue distribution is,
\be\label{sigma2} 
\sigma(\theta) = {1\over 2 \pi} (1 +   2\vr \cos\theta) \ \
\ \ 0 \leq 2\vr \leq 1 \ .
\ee
All these configurations also have free energy zero. 
\item
For $a_1>1$, there is a new saddle point whose eigenvalue distribution
function is given by
\be\label{sigma1}
\sigma(\theta) = {1 \over \pi \ \mt{\large sin}^2 \left ({\theta_0 \over
2} \right )} \sqrt{\mt{\large sin}^2 \left ({\theta_0 \over 2} \right
) - \mt{\large sin}^2\left ({\theta \over2}\right )} \mt{\large cos}
\left ({\theta \over 2} \right ) 
\ee 
where,
\be \label{theta0} 
\mt{\large sin}^2 \left ({\theta_0 \over 2} \right ) = 1 -
\sqrt{1 - {1 \over a_1(T)}} \equiv {1\over 2\vr}.
\ee
Note that this configuration is gapped unlike the above ones. 
The free energy for this configuration is (expressed in terms of $\vr$)
\be
F= -N^2 T \ltb \vr - {1\over 2} \ml(2 \vr) -  {1\over
2} \rtb \ .
\ee
\ei
Thus we see that there is a first order 
phase transition at $a_1=1$, which corresponds to a 
temperature $T=T_H$.

This was for zero coupling. Perturbatively, we have more complicated
matrix models Eq.\ref{weakacn}.  Restricting to models of the form
Eq.\ref{genZ}, such as the $(a,b)$ model, we can once again carry out
a saddle point analysis of the eigenvalue density, using a Hartree-Fock
approach. We now review the results (see \cite{spenta} for more
details)\footnote{One can also study the general models by expressing
them in terms of a suitable transform \cite{hong}, \cite{spenta},
\cite{spenta2} of the one plaquette matrix model
\cite{gw}\cite{wadia}.  This is particularly useful when studying the
vicinity of points where the large $N$ expansion breaks down.}.

The large $N$ saddlepoint equation for the model \ref{genZ} is 
\be
{-\hskip -12pt \int}\sigma(\theta) d\phi \mt{\large cot}\lb{\theta -
\phi \over 2} \rb = 2 S_{eff}^{\prime}(\sigma_1^2) \sigma_1 \mt{\large
sin}\ \theta 
\ee 
where $$\sigma_1={1\over N}{\rm Tr}U\ .$$ The solutions
of this saddle equation are given by, 
\be\label{revsol1} 
\sigma_1 =
\seff^{\prime}(\sigma_1^2) \sigma_1\ , {\hskip 2.2cm}0\le \sigma_1
\le {1\over 2} 
\ee and 
\be\label{revsol2} 
\seff^{\prime}(\sigma_1^2)
= {1 \over 4 \sigma_1 (1 - \sigma_1)}\ , {\hskip 1cm}{1\over 2} \le
\sigma_1 \le 1\ .  
\ee

The phase structure is as follows:
\bi
\item
(i) For sufficiently low temperature the only possible solution is of 
Eq.\ref{revsol1} 
\be
\sigma_1=0\ .  
\ee 

This stable saddlepoint in fact has the uniform distribution
Eq.\ref{sigunfrm}. It is identified with the thermal AdS saddlepoint
on the gravity side.
\item
(ii) At a higher temperature $T_0$, we find two new solutions, now of
Eq.\ref{revsol2} and thus with non-zero $\sigma_1$. The eigenvalue
distribution for both these distributions are of the same form as
Eq.\ref{sigma1} (with different values of the parameter $\vr$).  One
of these is stable and the other unstable.  On the gravity side, they
can be identified with the small black hole (SBH) and the big black
hole (BBH) respectively.
\item
(iii) There exists a temperature $T_1> T_0$ where the stable saddle
points of (i) and (ii) exchange dominance. This temperature
corresponds to the Hawking-Page temperature where the BBH has a lower
free energy than thermal AdS in the semi-classical gravity path
integral.
\item
(iv) At some temperature $T_c > T_0$, the eigenvalue distribution of
the unstable saddlepoint in (ii) changes from the gapped one in
\ref{sigma1} to the ungapped one in \ref{sigma2}. This
Gross-Witten-Wadia(GWW) like phase transition has been identified by
the authors of \cite{spenta} with the black-hole string transition
(see also \cite{azuma}).
\item
(v) And finally there exists a temperature $T_H$, the Hagedorn
temperature, when the unstable saddlepoint merges with the saddle
point in (i).  Above $T_H$ the saddlepoint in (i) becomes tachyonic.
\ei

It is quite remarkable how the general class of matrix models
\ref{genZ} captures all the detailed qualitative features of the
Hawking-Page phase diagram. This has been subsequently generalised to
the case where one has a charge or chemical potential \cite{basu}
\cite{yamada} (see also \cite{harmark}). One of the main new features
\cite{basu} argued in the case of fixed charge is that we have a term
in $S_{eff}(U)$ of Eq.\ref{sefftrunc} which is of the form $\ln((\mtr
U \mtr U^{\dagger})$. This has the effect that we no longer have the
saddle point (i) above with uniform distribution. This matches with
the known feature of the charged case that we never have thermal $AdS$
as a saddlepoint. The only saddlepoints are of the form \ref{sigma1}
and \ref{sigma2} and correspond to different kinds of black holes,
small and big, stable and unstable.  We refer the reader to
\cite{basu} \cite{yamada} for the details of the matching with the
gravity phase diagram.

\section{Exact Solution at Finite $N$} \label{secexatsol}

In this section we will obtain an exact expression for the gauge
theory partition function \ref{freepf} of the free theory. As we will
see the method of solution can be straightforwardly generalised to the
general case described by Eq.\ref{weakacn}.

Starting with the matrix model which captures the free gauge theory,
\be\label{Zg}
Z = \int [d U]\exp \left[\sum \limits_{n=1}^{\infty} {a_n(T) \over
    n}  \mtr U^n \mtr U^{\dagger n} \right]\ ,
\ee
we can expand the exponential to obtain for the integrand
\be\label{frob}
\exp \left[\sum \limits_{n=1}^{\infty} {a_n(T) \over n}  \mtr U^n  \mtr
  U^{\dagger n} \right]= \sum \limits_{\vec k} {1 \over z_{\vec k}}
\prod_{j}a_j^{k_j}\mU_{\vec k} (U) \mU_{\vec k} (U^{\dagger}).
\ee
Here,
\be
z_{\vec k} = \prod \limits_j k_j! j^{k_j}
\ee
and
\be
\mU_{\vec k}(U) = \prod \limits_{j=1}^{\infty} ( \mtr U^j )^{k_j} .
\ee

It is convenient to write this in terms of group characters using the
Frobenius formula,
\be
\mU_{\vec k}(U) = \sum \limits_R \chi_R(C(\vec k)) \ \mtr_R U ,
\ee
where $\chi_{R}(C(\vec k))$ is the character of the conjugacy class
$C(\vec k)$ of the permutation group\footnote{Recall that a conjugacy
class of the permutation group can be labeled by a partition $\vec k
=(k_1,k_2,\ldots)$. $\vec k$ is an infinite dimensional vector with
$k_j$ being the number of cycles of length $j$.} $S_K$, ($K= \sum j
k_j$), in the representation $R$ of $U(N)$.

Now we can carry out the integral over the holonomy
using the 
orthogonality relation between the characters of  \un \footnote{
The invariant Haar measure $[dU]$ that appears here has been normalized
such that $\int [dU] =1$.},
\be\label{orthonor2} 
\int [d U] \ \mtr_R (U) \ \mtr_{R'}(U^{\dagger}) = \delta_{R R'}\ .
\ee 
Therefore we obtain 
\be \label{zexact}
Z(\beta) = \sum_{\vec k}{\prod_{j}a_j^{k_j} \over z_{\vec k}} \sum_{R} 
\left[\chi_R(C(\vec k)) \right]^2\ .
\ee
This is an exact expression for any $\beta$ and $N$ for the partition
function of the free gauge theory.  In the next section we will
further analyse the properties of this solution.  However, it should
be noted that the answer is completely explicit. The sum over
representations of $U(N)$ can be labelled by Young Tableaux with $N$
rows and arbitrary numbers of boxes in each row.  The characters of
the conjugacy class are determined recursively by the Frobenius
formula \cite{hamermesh}. Explicit expressions for the most general
case are not simple but have been given in the literature
\cite{lasalle}.

In the strict $N=\infty$ limit where the sum over representations is
unrestricted, we can use the group theory identity for the
orthogonality of characters of different conjugacy classes (see for
instance, \cite{hamermesh} pg.110) to obtain 
\be\label{zinfty}
Z(\beta)=\sum_{\vec k}\prod_{j}a_j^{k_j}= \prod_j(1-a_j)^{-1}\ .  
\ee
This agrees with the exact $N=\infty$ answer derived in \cite{shiraz}.

In the special case where only $a_1\neq 0$, the exact answer
\ref{zexact} is given by the simpler expression, 
\be\label{za1}
Z(\beta) = \sum \limits_{k=0}^{\infty} \sum \limits_R {1 \over k!}
\left[d_R(S_k) \right]^2 a_1^{k}\ .  
\ee 
In this case the only conjugacy
class that contributes is the trivial or identity class consisting of
$k$ one cycles. The character of this class is nothing but the
dimension $d_R(S_k)$ of the representation $R$ for the permutation
group $S_k$.

It is clear that the above method of solution can be straightforwardly
generalised to the matrix models \ref{weakacn} which describe the
perturbative gauge theory at finite temperature.  We can similarly
expand the exponential and carry out the unitary integrals after using
the Frobenius relations. The general answer for the finite $N$ matrix
model can once again be explicitly written though the actual
expressions will now be more cumbersome. As a special case consider
the $(a,b)$ model \ref{Zab}.  
\be\label{Zab2} Z(a_1,b_1) = \int [dU] \exp
\ltb a_1\mtr U \mtr U^{\dagger} + {b_1 \over N^2} \lb \mtr U \mtr
U^{\dagger} \rb^2 \rtb \ .
\ee 
Expanding the exponential as before and
using a similar logic as before, we can write the \pf as,
\be\label{Zabans} Z(a_1,b_1) =\sum \limits_{k=0}^{\infty} \sum
\limits_{l=0}^{k/2} {a_1^{k-2l} b_1^{l} k! \over N^{2l} l!  (k-2l)!}
\sum_R {d_R^2 (S_k) \over k!} \ .  
\ee

For the more general case of models Eq.\ref{genZ} with
\be\label{sefftrunc2} 
\seff(U)=\sum_{k=1}^{\infty}{\alpha_k\over
N^{2(k-1)}}(\mtr U \mtr U^{\dagger})^k \ .  
\ee 
We can again go through
the same steps to obtain 
\be\label{genZans} 
Z(\{ \alpha_i\}) = \sum
\limits_{\{k_l=0\}}^{\infty} \prod_l\lb {\alpha_l^{k_l} \over
N^{2k_l(l-1)}k_l!} \rb \sum_R d_R^2(S_K) \ ;\ \ \ \ \ \ K= \sum
\limits_{l=1} l k_l .  
\ee 
These are the cases whose large $N$ limit
we will be analysing in some detail in what follows.

Finally, we should remark that matrix models of the form Eq.\ref{Zg}
also appear in the counting of BPS states \cite{raju}. In fact, it is
not difficult to use certain standard identities for the completeness
of group characters to evaluate the answer Eq.\ref{zexact} in the case
where $a_n=a_1^n$. One reproduces the usual generating function for
the half BPS states given in \cite{raju}. It would be interesting to
use the exact answer for finite $N$ to evaluate some of the partition
functions/indices of interest in ${\cal N}=4$ Super Yang-Mills 
theory \cite{raju}.


\section{Taking the large $N$ Limit} \label{seclargeN}

We start by analysing the large $N$ behaviour of the exact answer (for
the free gauge theory) given in \ref{zexact}\footnote{We will
generalise to the interacting case in Sec.\ref{secnonzerocup}}.  We
should be able to see the large $N$ phase transition that was obtained
from the analysis of the eigenvalue density (reviewed in
Sec.\ref{secrev}). We can see, in general, from the form of the
solution that, in the large $N$ limit, there is likely to be a
dominant representation contributing in the sum over
representations. Essentially, this can be viewed as a statistical
mechanical system in which the group characters behave like an entropy
contribution (roughly favouring representations with a large number of
boxes in the \yt \ ). And the $a_i$ are the Boltzmann suppression
factors which disfavour representations with a large number of
boxes. The balance between them leads to a dominant representation at
any particular value of the temperature.  A large $N$ phase transition
would occur when the nature of this dominant representation undergoes
a qualitative change as one varies the temperature.  As mentioned
earlier, the Douglas-Kazakov \cite{dk}\cite{gm} phase transition and
its generalisations in $2d$ Yang-Mills theory can be understood this
way.  We will now see all these features explicitly in our matrix
models, specialising for simplicity to the special case where only
$a_1\neq 0$.  As mentioned in Sec. \ref{secevana}, this case captures
all the essential physics of the finite temperature theory.

In the special case where 
$a_n =0$ \ for \ $n > 1$, the exact answer is given by Eq.\ref{za1}  
\be\label{za11}
Z(\beta) = \sum \limits_{k=0}^{\infty} \sum \limits_R {1
\over k!} \left[d_R(S_k) \right]^2 a_1^{k} \ .
\ee
To proceed, we will write the sum over representations of $U(N)$
in terms of the number of boxes of the corresponding \yt \ 
\be 
\sum \limits_R \rightarrow \sum \limits_{k=1}^{\infty} \sum
\limits_{\{n_i\}=0}^{\infty} \ \delta \left ( \Sigma_{i=1}^N n_i - k
\right ) \ \ \mt{\large with} \ \ n_1 \ge n_2 \ge . . .\ge n_N \ge 0\ ,
\ee 
where $n_i$ is the number of boxes in the $i^{th}$ row of \yt \ (there
being only $N$ rows for a representation of $U(N)$). Also $k$ is the
total number of boxes in the representation. Therefore the \pf reads
as
\be\label{za12}
Z(\beta)=\sum \limits_{k=0}^{\infty}\sum \limits_{\{n_i\}= 0}^{\infty} 
{1 \over
k! } \left[d_R(S_k) \right]^2 a_1^{k} \
\delta(\Sigma_{i=1}^N n_i - k )\ .
\ee 
The dimension $d_R(S_k)$ is given by the Frobenius-Weyl 
formula \cite{hamermesh}
\be\label{dim}
d_R(S_k) = {k! \over h_1! h_2! ... h_N!} \prod \limits_{i < j}
(h_i -h_j)\ ,
\ee
where,
\be \label{h-n}
h_i = n_i +N-i\ ,
\ee
with \be
h_1>h_2>...>h_N \geq 0\ .
\ee
{\vskip 1cm}

\subsection{The Continuum Limit and Saddlepoint Equations} 
\label{secconti}

In the $N \ra \infty$ limit we can define, following Douglas and
Kazakov \cite{dk}, continuous functions which describes each young
tableaux 
\be n(x) = {n_i \over N}, \ \ h(x) = {h_i \over N}\ , \ \ \ \
x={i \over N}\ , 
\ee 
where $x \in [0,1]$.  The function $n(x)$ or
equivalently $h(x)$ captures the profile of the large $N$ Young
tableaux.  In this limit Eq. \ref{h-n} can be written as, 
\be
\label{hx-nx} h(x) = n(x) + 1 -x\ .  
\ee 
Note that the condition $n_1
\ge n_2 \ge .. \ge n_N$ implies a strict monotonicity for $h(x)$ 
\be
\label{h-cond} h(x) > h(y) \ \ \ \mt{\large for} \ \ \ y > x.  
\ee
In this limit the total number of boxes in a \yt \ is given by,
\be \label{k-k'}
k =\sum_{i=1}^N n_i \ \ra \ N^2 \left(\int_0^1 dx \ [h(x)
+x -1] \right) = N^2 \left(\int_0^1 dx h(x) -{1 \over2}\right)\ .
\ee
Since $k' \equiv \int_0^1 dx h(x) -{1\over 2}$ will generically be
${\cal O}(1)$, we see that the number of boxes $k=N^2 k'$, in a
generic representation, is of the order of $N^2$ in this limit.

The \pf Eq.\ref{za12} can be written, using Eq. \ref{dim}, as
\be\label{Za13}
Z(\beta) = \sum_{\{h_i\}} \exp \lb \ln(k!) +\ln a_1 +
\sum_{i \neq j} \ln |h_i-h_j| - 2\sum_i \ln (h_i!) \rb. 
\ee
In the large $N$ limit, using Stirling's approximation for the
factorials and Eq.\ref{k-k'}, the partition function can be expressed
as
\be\label{zcont}
Z  = \int [dh(x)] \me^{- N^2 \seff},
\ee
where,
\ben\label{Seff}
- \seff &=& \int_0^1 dx \ {-\hskip -11.5pt}\int_0^1 dy \ln
 |h(x)-h(y) | \nn &-& 2\int_0^1 dx h(x) \ln h(x)
+k' \ln\lb{a_1k'}\rb +k'+1\ .  
\een
Recall that  $k' \equiv \int_0^1 dx h(x) -{1\over 2}$.

Now we are in a position to carry out a saddlepoint analysis 
for the effective action functional $\seff[h(x)]$  
(\ref{Seff}). Varying $\seff$ with respect to $h(x)$, we obtain the
saddlepoint equation,
\be\label{sadeq1}
{-\hskip -12pt}\int_{0}^1 {dy \over h(x) - h(y)} = \ln h(x)
-  {1 \over 2}\ml \ltb a_1k' \rtb. 
\ee

Introduce, again following \cite{dk}, the density of boxes in
the \yt \  $u(h)$ defined by
\be\label{u}
u(h) =  - {\p x(h) \over \p h}.
\ee
By definition, it obeys the normalisation
\be \label{unorm}
\int_{h_L}^{h_U} dh u(h) = 1,
\ee
where the interval of support $[h_L, h_U]$ of $u(h)$ is specified by
$h_L = h(1)$ and $h_U = h(0)$.  From the monotonicity of $h(x)$
Eq.\ref{h-cond}, it follows that $u(h)$ obeys the constraint
\be\label{uconstr} u(h)\leq 1\ .  
\ee

In terms of the density $u(h)$, the saddle-equation (\ref{sadeq1})
can be written in the more familiar form,
\be \label{sadeq2}
{-\hskip -12pt}\int_{h_L}^{h_U} dh' {u(h') \over h  -  h'} = 
\ln [h] - {1 \over 2} \ln \ltb a_1k' \rtb = \ln\ltb{h\over \vr}\rtb
\ee
where $\vr^2\equiv a_1 k'$. 
Note that the parameter $\vr$ involves $k'$ given by 
\be\label{kprime}
k'=\int_0^1 dx h(x) -{1\over 2}=\int hu(h)dh -{1\over 2}
\ee
which in turns depends on the (first moment of the) density $u(h)$. We
will therefore have to solve the equation self-consistently.

\subsection{The Saddlepoint Densities} \label{secsadden}

\begin{figure}
\centering
\includegraphics[height=6cm]{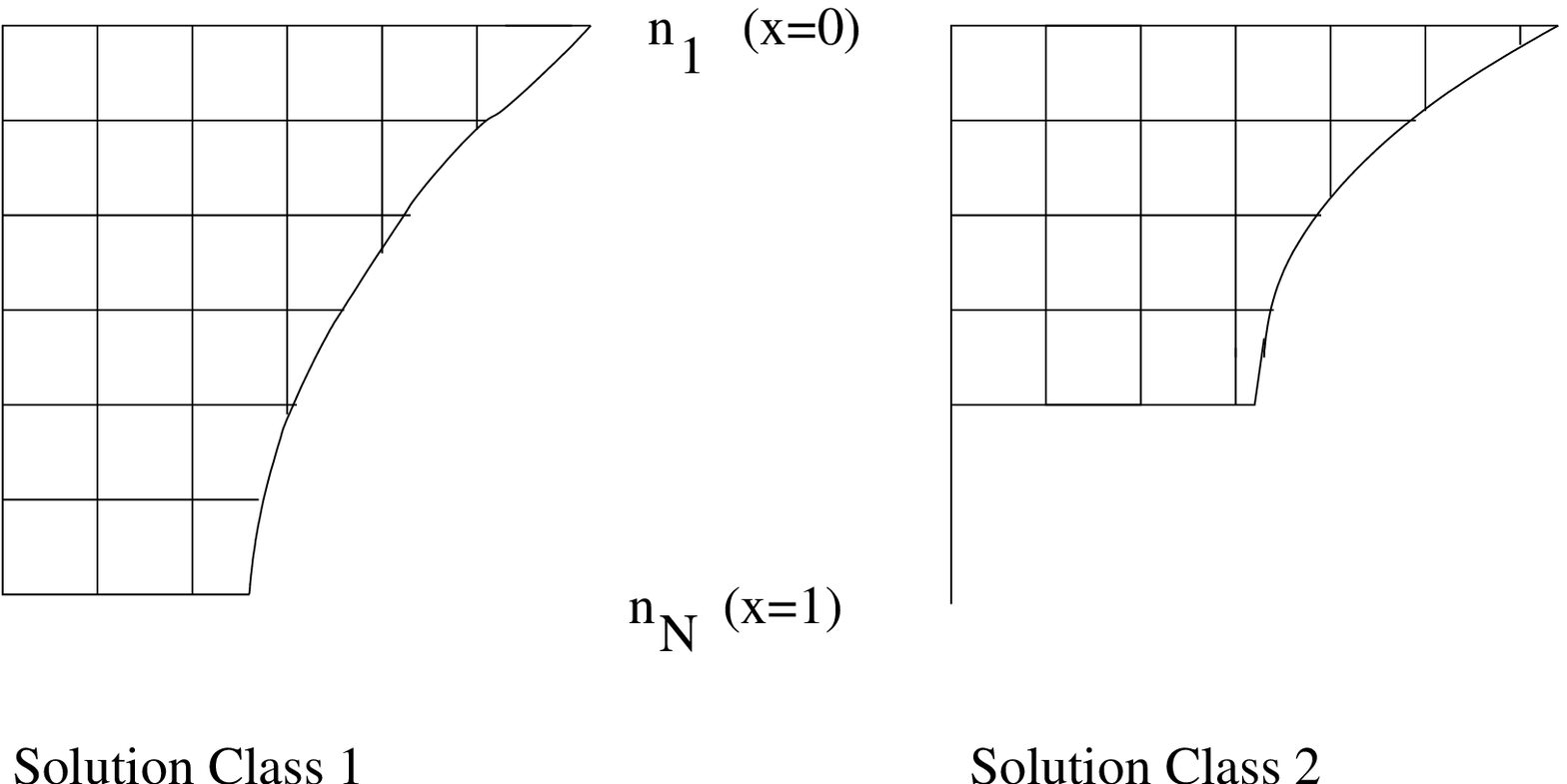}
\caption{\yt}
\label{ytrep}
\end{figure}

The solution to the integral equation \ref{sadeq2} for the \yt\
density $u(h)$ is obtained along similar lines to the usual solution
for the eigenvalue density.  The main point to additionally take into
account is the presence of the constraint Eq.\ref{uconstr}.  Thus we
will find that the solutions to \ref{sadeq2} are of two different
kinds depending on the value of the parameter $\vr$. \\ \\
{\bf Solution Class 1:}
\be\label{bra1}
0 \leq u(h) <1; \ \ \ \ \  h \in [\hm, \hma]\ . 
\ee
A typical representation corresponding to such a \yt\ density  is
plotted in fig\ref{ytrep}. There are always a nonzero number of boxes in
each row. \\
\\
{\bf Solution Class 2:}
\ben \label{bra2}
u(h) &=& \ 1  {\hskip 1cm} h \in [0,\hm] \nn
&=& \tilde u(h) {\hskip .55cm}  h \in [\hm, \hma]
\een
with $0\leq \tilde u(h) <1$.
A typical young tableau for a solution of this class 
has been plotted in Fig.\ref{ytrep}. The representations are such that 
a finite fraction of the rows are empty. 

As we vary $\vr$, the constraint \ref{uconstr} will come into play and
one will have to switch from one of the branches to the other. At this
point, as we will see explicitly, there will be a non-analyticity, for
example in the free energy and we will have a large $N$ phase
transition.

We will now solve the saddle-equation in the conventional 
way\footnote{For a recent review see \cite{marino}.} by introducing the
resolvent $H(h)$ defined by 
\be \label{H} H(h) = \int_{h_L}^{h_U} dh'
{u(h') \over h - h'}\ .  
\ee 
The resolvent has the following
properties, 
\bi 
\ii (i) \ It is analytic in the complex $h$ plane with
a branch cut along the positive real interval $(h_L,h_U)$.
\ii (ii) \ It is real for real
positive $h$ outside the interval.  
\ii (iii) \ $H(h) \sim {1 \over h}
+ (k' + {1 \over 2}){1 \over h^2}$ \ \ ({\rm as} $h\ra \infty$). This
follows from the moment expansion of the resolvent at large $h$ and
using Eq.\ref{kprime}.  
\ii (iv) \ $H(h+ i \epsilon) + H(h- i
\epsilon) = 2\ml \ltb {h \over \vr} \rtb$ for real $h$ .  \ii (v) \ \
$u(h) = - {1 \over 2 \pi i} \ltb H(h+ i \epsilon) - H(h- i \epsilon)
\rtb$ for $h\in [\hm, \hma]$.  \ei

One can therefore solve for $H(h)$ in terms of its real part by
writing it as a contour integral. In fact, the equations we need to
solve are very close to the equations that arise in a class of matrix
models studied by Kazakov, Staudacher and Wynter\cite{ksw} (see also
\cite{arsiwalla} for the solution of a similar equation).  We now
exhibit this solution in both the classes mentioned above.

\subsubsection{Solution Class 1:} \label{secsol1}

In branch $1$ using the ansatz \ref{bra1} the saddle equation becomes,
\be \label{sadeqbra1}
{-\hskip -12pt}\int_{\hm}^{\hma} dh' {u(h') \over h - h'} = \ml \ltb
{h \over \vr} \rtb \ ,{\hskip .5cm} h \in [\hm\ , \hma].
\ee
The resolvent (Eq. \ref{H}) in this branch is given by (see
\cite{ksw},\cite{marino} for a general discussion),
\be \label{Hbra1}
H(h) =  - \sqrt{(h - \hma) \ (h - \hm)} \oint {ds \over 2 \pi i}
{\ml \ \lb s/\vr \rb \over (s  -  h) \sqrt{(s - \hma) \ (s  - 
\hm)}}\ .
\ee
\begin{figure}
\centering
\includegraphics[height=6cm]{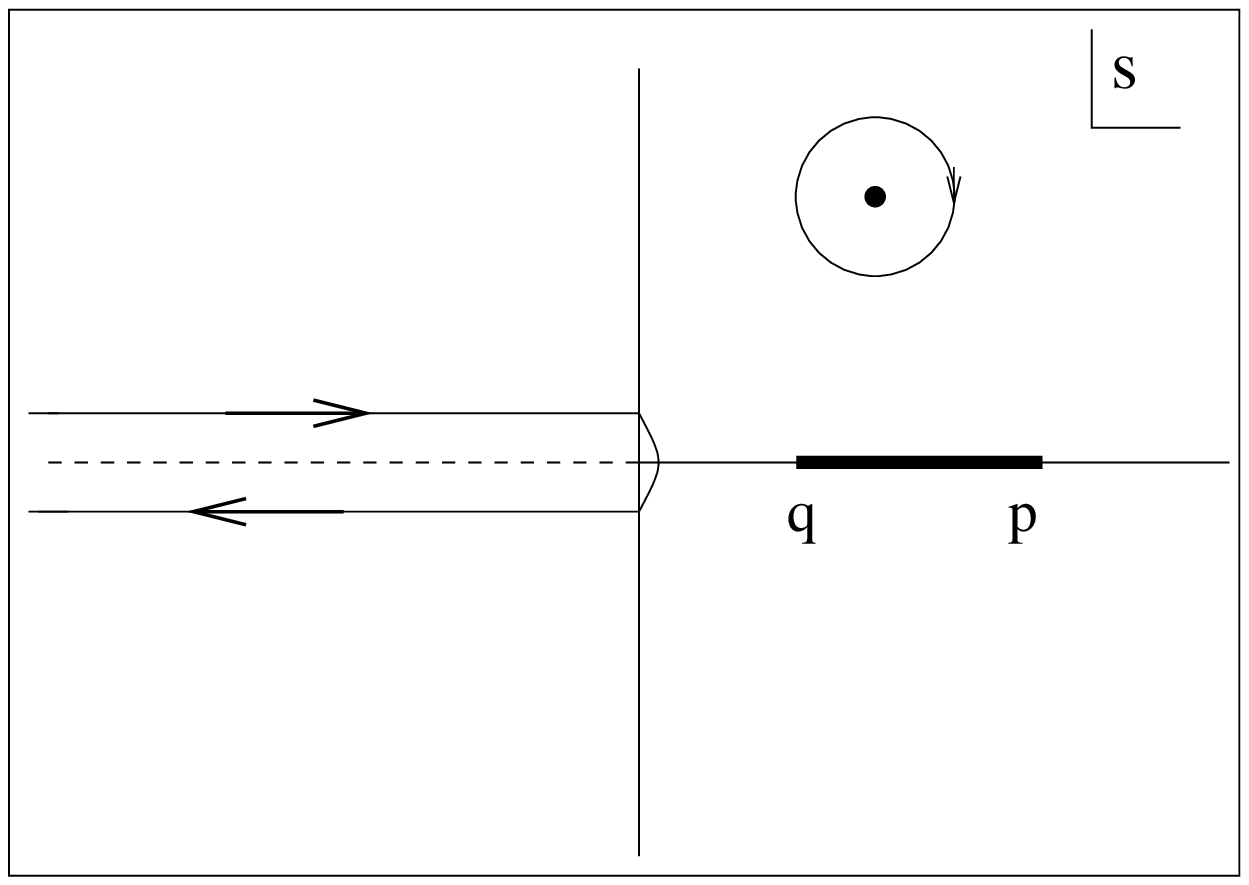}
\caption{Contour for Solution class 1}
\label{cont_b1}
\end{figure}
The contour of integration is shown in Fig.\ref{cont_b1}. Carrying out
the contour integration we obtain the resolvent for class 1,
\be \label{Hsol1}
H(h) = \ml \ltb {2 \ h^2 - (\sqrt{\hma} - \sqrt{\hm})^2 \ h + 2 \ \hm
\ \hma - 2 \ (h + \sqrt{\hm \ \hma}) \sqrt{ (h- \hma) (h - \hm)} \over
\vr \ (\sqrt{\hma} + \sqrt{\hm}) ^2 } \rtb \ .
\ee

\begin{figure}
\centering
\includegraphics[height=6cm]{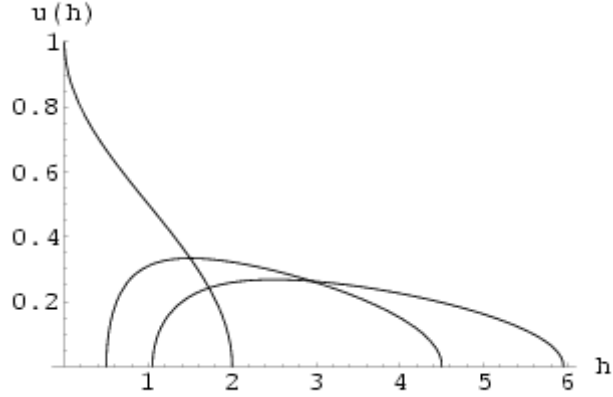}
\caption{Plot of $u(h)$ vs. h for Solution Class 1. The value of $\vr$
increases from 0.5 as one goes from the leftmost graph to the right.}
\label{hfig1}
\end{figure}

Using this, we can readily find the discontinuity and thus the \yt \
density function $u(h)$,
\ben \label{usol1} u(h) &=& {1 \over \pi} \mt{\large cos}^{-1} \ltb
{h-1 \over 2\ \vr} + {\lb \vr - {1 \over 2} \rb^2 \over 2 \ \vr \ h}
\rtb \ \nn 
&& {\hskip 6.5cm} \mt{\large for} \ h \in [\hm, \hma] \nn
&=& {2 \over \pi}\mt{\large cos}^{-1} \ltb {h + \vr - 1/2 \over 2
\sqrt{ \vr h} }\rtb.  \een
Fig.\ref{hfig1} shows the plot of $u(h)$ vs. h for this solution class.

The support of $u(h)$ as well as $k'$ is determined by expanding
$H(h)$ for large $h$ and matching with property (iii) of $H(h)$ listed
above.
\be
H(h\ra \infty) \sim \ml \ltb {(\sqrt{\hm} + \sqrt{\hma})^2 \over 4 \
\vr} \rtb + {(\sqrt{\hma}- \sqrt{\hm})^2 \over 2} {1 \over h} + \lb
\sqrt{\hm \hma} + {3 \over 4} \rb {1 \over h^2}\ .
\ee
We therefore obtain
\ben\label{pqval}
\sqrt{\hm} &=& \sqrt{\vr}  - {1 \over \sqrt{2}}, \\
\sqrt{\hma} &=&  \sqrt{\vr}  + {1 \over \sqrt{2}} 
\een
and 
\be \label{a1-vr}
k' = \sqrt{\hm \hma}  +  {1 \over 4} 
\ee
which implies 
\be \label{a1vr}
a_1 = {4  \vr^2 \over 4  \vr -1}
\ee
using the definition $\vr^2=a_1k'$.  Since $\hm$ is a real positive
quantity, Eq.\ref{pqval} implies that this solution branch exists for
$\vr \ge {1 \over 2}$. From Eq.\ref{a1vr}, we therefore conclude that
this class of solutions only exist for $a_1\geq 1$.

\subsubsection{Solution Class  2:} \label{secsol2}

\begin{figure}
\centering
\includegraphics[height=6cm]{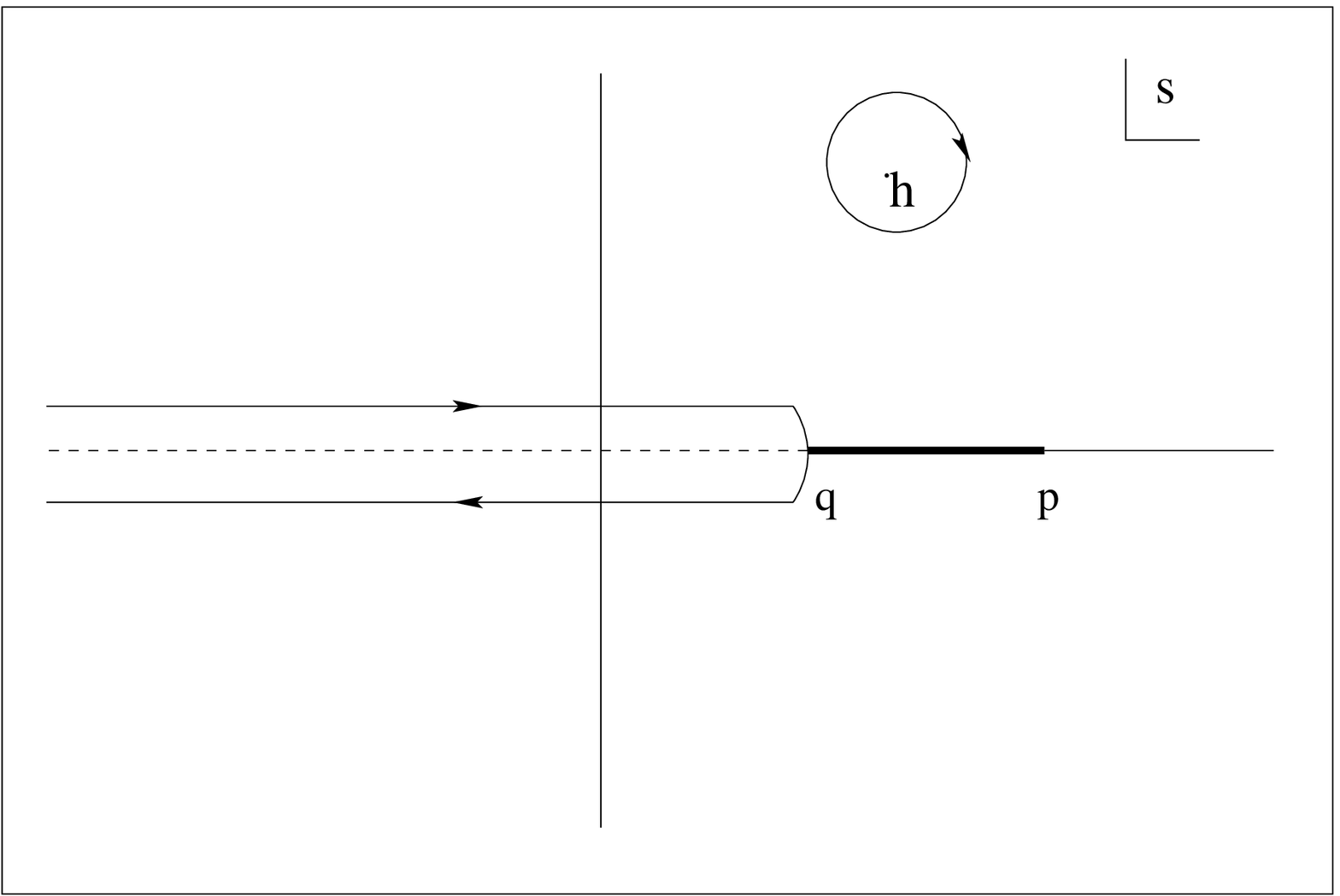}
\caption{Contour for Solution Class 2}
\label{cont_b2}
\end{figure}

\begin{figure}
\centering
\includegraphics[height=6cm]{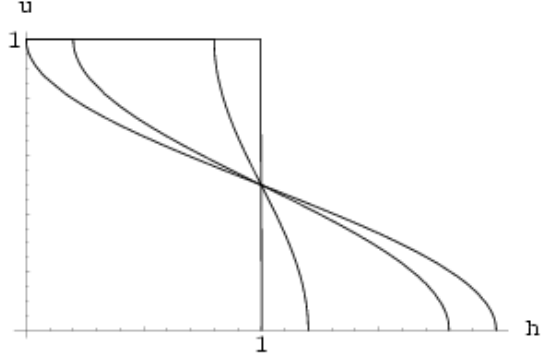}
\caption{Plot of $\tilde u(h)$ vs. h for Solution Class 2. The value
of $\vr$ decrease from 0.5 to 0 as one goes from the rightmost to the
leftmost graph.}
\label{hfig2}
\end{figure}

Using the ansatz \ref{bra2} the saddle equation becomes,
\be \label{sadeqbra2}
 {-\hskip -12pt}\int_{\hm}^{\hma} dh' {\tilde u(h') \over h - h'}
= \ \ml \ltb {h \over \vr} \rtb - \ml \ltb {h \over h- \hm } \rtb
{\hskip .5cm} \mt{\large where,} \ h \in [\hm, \hma] \ .
\ee
The full resolvent $H(h)$ now takes  the form,
\be
H(h) = \ml \ltb {h \over h-\hm} \rtb + \int_{\hm}^{\hma} dh'  {\tilde
u(h') \over h  -  h'} \ .
\ee 
$H(h)$ can once again be written as a contour integral,
\be\label{Hbra2}
H(h) = \ml \ltb {h \over h-\hm} \rtb - \sqrt{(h - \hma) \ (h - \hm)}
\oint {ds \over 2 \pi i} { \ml \lb s/\vr \rb - \ml \ltb s/( s
- \hm)\rtb \over (s - h) \sqrt{(s - \hma) \ (s - \hm)}}\ .
\ee
The contour is shown in fig \ref{cont_b2}. 
Carrying out the integration gives the answer
\be \label{Hsol2}
H(h) = \ml \ltb {h \over \vr} \rtb + \ml \ltb {\sqrt{\hma} -
\sqrt{\hm} \over\sqrt{\hma} + \sqrt{\hm} } \rtb + \ml \ltb {h + \sqrt{\hm 
\hma} - \sqrt{ (h-1)^2 - 4 \vr^2} \over h - \sqrt{\hm  \hma} + \sqrt{
(h-1)^2 - 4 \vr^2}} \rtb \ .
\ee
Hence, the \yt \ density is given by,
\be \label{usol2}
\tilde u(h) =  {1 \over \pi}  \mt{\large cos}^{-1} \ltb {h-1 \over 2
\vr} \rtb \ .
\ee
Fig.\ref{hfig2} shows the plot of $\tilde u(h)$ vs. h for this solution class.

As before, expanding $H(h)$ for large $h$ we find the values of
$\hm$, $\hma$ and $k'$ as follows,
\ben \label{pqva2}
\hm &=& 1 -  2\vr \ , \\
\hma &=& 1  + 2\vr 
\een
and 
\be\label{k-vr2}
k' = \vr^2 \ .
\ee
From the definition $a_1k'=\vr^2$ we obtain
\ben\label{a1-vr2}
\mt{\large either} \ \vr &=&0 \nn
\mt{\large or}, a_1 &=&  1.
\een
The former implies the uniform distribution
\be\label{unfrm}
u(h)=1 \ \ \ \ h \in [0,1].
\ee
This is therefore a saddlepoint for any value of $a_1$. This is in
fact the density corresponding to the trivial representation $n_i=0$.

The latter corresponds to a family of saddlepoints labelled by $\vr$
which exists only at $a_1=1$. From Eq. (\ref{pqva2}) it is clear that
this family exists for $\vr \le 1/2$.

To summarise, we see that there are three different saddlepoint
configurations of Young Tableau densities.  
\bi
\item
The trivial representation corresponding to the uniform distribution
for $u(h)$, Eq. \ref{unfrm}.  This exists for any value of $a_1$,
i.e. for any temperature.
\item
The continuous family Eq. \ref{usol2} which exists only for $a_1=1$. Here
a finite fraction of the rows of the Young Tableau are empty.
\item
The representation Eq. \ref{usol1} which exists only for $a_1>1$,
i.e. only for high enough temperature.  Now all the $N$ rows of the
\yt\ are filled.  
\ei

We see that the saddlepoints are exactly in correspondence with the
saddlepoints of the eigenvalue density, as summarised in
Sec.\ref{secevana}. We will now indeed verify that the large $N$ free
energy of these saddle points is also exactly the same as that seen
from the eigenvalue density analysis.


\subsection{The Free Energy of the Free Theory} \label{secfreeen}

In the large $N$ limit the free energy of the partition function
Eq. \ref{zcont} is given by,
\be
F= - T\ml Z \ = N^2 T \seff^0 \ .
\ee
where $\seff^0$ is the value of effective action at the (dominant)
saddlepoint.  The effective action given in \ref{Seff} can be
re-expressed as a functional of $u(h)$ 
\be\label{Seffyt} -\seff =
\int_{h_L}^{h_U} dh {-\hskip -12pt}\int_{h_L}^{h_U} dh' u(h) u(h') \ml
|h - h' | - 2\int_{h_L}^{h_U} dh \ u(h) \ h \ \ln[h] +k' +
1 + k' \ \ml [a_1k']. 
\ee 
We only need to evaluate this functional on the
different saddlepoint configurations we have found.  In evaluating
the expressions it is useful to use the corresponding saddlepoint
equations to eliminate the quadratic term in $u(h)$ in the effective
action. We thus obtain: 
\bi
\item
{\bf Solution Class 1:}
Using the saddle equation Eq. (\ref{sadeqbra1}), $\seff$ becomes
\be \label{Eeffonshel1}
-\seff^0 = - \int_{\hm}^{\hma} dh u(h) h \ml h + \lb 
\vr - {3\over 4} \rb
\ml [\vr]  + {1 \over 2} + {\cal C}_1 \ ,
\ee
where the constant ${\cal C}_1$ is given by,
\be 
\label{C1} {\cal C}_1 =  {1 \over 2} \ml[\vr/2] + \vr - {1\over
2}.  \ee 
Evaluating the integral (details can be found in Appendix A)
gives finally for the effective action, 
\be \label{F1} F = N^2 T
\seff^0 =- N^2 T \ltb \vr - {1\over 2} 
\ml(2 \vr) - {1\over 2} \rtb
\leq 0.  \ee 
This exactly agrees with the free energy computed in
\cite{shiraz} for the deconfined phase saddlepoint which was quoted
in Sec.\ref{secevana}.

\item
{\bf Solution class 2 :}

Here the free energy is easy to compute since we have a continuous
family labelled by $\vr$ which are all saddlepoints and thus must
have the same free energy when $a_1=1$.  In particular the constant
configuration Eq.\ref{unfrm} is a limiting member of this family for
which the effective action is readily computed to be zero. Therefore
the entire family of saddlepoints in Eq.\ref{usol2} must have zero
free energy. This is explicitly verified in appendix A.  Once again
this matches with the results of the eigenvalue analysis for the
confined phase saddlepoint.  \ei
 
We see from this that there is an exchange of dominance of the saddle
points at $a_1\geq 1$, where one gets a new saddlepoint
Eq.\ref{usol1}, which has less free energy compared to the uniform
density saddlepoint that exists for all values of $a_1$.  Therefore as
the temperature increases we get a transition when the saddlepoint
switches giving rise to the confinement-deconfinement transition. We
see that this is nothing bu the analogue of the Douglas-Kazakov
transition in our approach. We have thus reproduced the usual
saddlepoints as well as the phase diagram of the zero coupling
Yang-Mills theory from our method of taking the large $N$ limit of the
exact solution.


\subsection{Extension to Non-zero Coupling} \label{secnonzerocup}

The general interacting unitary matrix model Eq.\ref{weakacn} for
perturbative gauge theory can also be solved exactly using the
technique of Sec.\ref{secexatsol} and a large $N$ limit can then be
taken along the lines described in this section. However, the analysis
is going to be technically much more involved.  But as described in
Sec.\ref{secrev}, the essentials of the physics is in any case
captured by models involving only ${\rm Tr}U$. In particular, the
$(a,b)$ model Eq.\ref{Zab} already does a good job in getting the
detailed form of the Hawking-Page phase diagram\cite{spenta}. Here we
will sketch how its exact solution Eq.\ref{Zabans} shows all the
features described in Sec.\ref{secevana} when we take the large $N$
limit as described in this section.

Let's write the answer Eq.\ref{Zabans} as
\ben\label{Zab3} 
Z(a_1,b_1) &=& \sum \limits_{k=0}^{\infty} \sum
\limits_{l=0}^{k/2} {a_1^{k-2l}  b_1^{l}  k! \over N^{2l}  l! 
(k-2l)!} \sum \limits_{\{n_i\}} {d_{\{n_i\}}^2 (S_k) \over k!} 
\delta(k-\Sigma_i n_i) \nn 
&=& \sum \limits_{k=0}^{\infty} f(k) 
\sum \limits_{\{n_i\}} {d_{\{n_i\}}^2 (S_k) \over k!} 
\delta(k-\Sigma_i n_i) \ ,
\een
where, as usual, $\{n_i\}$ label the number of boxes in the \yt\ and  
\be\label{sumf}
f(k) = \sum \limits_{l=0}^{k/2}  {a_1^{k-2l}  b_1^{l} k! \over
 N^{2l}  l!  (k-2l)!}.
\ee
Since the total number of boxes $k$ is of order $N^2$, we can replace the sum 
in Eq. \ref{sumf} by the saddlepoint value.  Doing this gives
\be\label{sadf}
f(k)={k\over 2}\lb{a_1 \over 1-x}\rb^ke^{-{kx\over 2}}\equiv {k\over
2}{\tilde a_1}^k
\ee
where $x$ is determined by the equation 
\be
{x \over (1-x)^2}={2 b_1 k'
\over a_1^2}
\ee
 with $k=N^2k'$.  Therefore the \pf
\be 
Z(a_1,b_1) = \sum \limits_{\{n_i\}} {k \over
2}{\tilde a_1}^k \ {d_{\{n_i\}}^2 (S_k) \over k!} , 
\ee 
(with
$k=\sum_i n_i$) takes essentially the same form as Eq.\ref{za12}
except that we have ${\tilde a_1}$ instead of $a_1$.\footnote{The extra
multiplicative factor of ${k\over 2}$ plays only a subleading role in
the large $N$ limit.}

We can now take the large $N$ limit as before to obtain
\be\label{Zabcont} Z(a_1,b_1) = \int [dh(x)] \ \me^{- N^2 \seff}, 
\ee
where $\seff$ is the same as in Eq.\ref{Seff} with the replacement of
$a_1$ by ${\tilde a_1}$.

Since ${\tilde a_1}$ depends on $k'$ or $h(x)$ (see \ref{sadf} and
below), the saddle equations are modified a bit. Taking into account
this additional dependence on $h(x)$, the saddle equation becomes,
\be\label{sadeqab} {-\hskip -12pt}\int_{h_L}^{h_U} dh' {u(h') \over h
- h'}= \ml \ltb h(x) \over \tx \rtb \ \ \mt{\large where} \ \ \tx^2 =
\tilde a_1 k' \me^{x\over 2}.  
\ee

Since the saddle equations are of the same form as Eq.\ref{sadeq2},
(with the replacement of $\vr$ by $\tx$) the saddlepoint
configurations for the \yt\ density are also the same in form. Namely,
we obtain the three different configurations of
Sec.\ref{secsadden}. In fact, we can redo this analysis for the class
of models Eq.\ref{genZ} for which the exact solution was given in
Eq.\ref{genZans}. This is performed in Appendix \ref{appB}. We see
from the analysis there that the saddlepoint equations give once again
the same saddlepoint configurations.  Moreover, we obtain the same
Hartree-Fock equations as Eq. \ref{revsol1} and
Eq. \ref{revsol2}. Thus the phase diagram turns out to be the same as
that given by the eigenvalue density analysis. For instance in the
case of the $(a,b)$ model we find
\bi
\item
A low temperature saddlepoint which is characterised by $\tx=0$ which
is the uniform distribution corresponding to thermal $AdS$. This has
zero free energy.
\item
Then there is a saddlepoint of the form Eq. \ref{usol2} when $\tx^2 = k'$
i.e. ${\tilde a_1}e^{x\over 2}=1$ and $\tx \le {1\over 2}$.  This
implies that 
\be \tilde{\xi}^2 = {1-a_1 \over 2 b_1} \le {1 \over 4}\
.  \ee 
This is actually the unstable saddlepoint corresponding to the
small black hole (in the phase where it is to be viewed as an excited
string state) and has positive free energy \be F = N^2T {(1-a_1)^2
\over 4 b_1}\ .  \ee This saddlepoint exists in a temperature range
$T_c \le T \le T_H$.
\item
Finally there is the saddlepoint of the form Eq. \ref{usol1} which 
obeys
\ben
{a_1 \over 1-x} &=& {4 \tx^2 \over 4\tx - 1} \nn
\mt{\large and} \ \ k' &=&  \tx -{1 \over 4}\ .
\een  
The two solutions to this equation give rise to the BBH as well as the
SBH (in the actual black hole regime).  The BBH solution exists for
all temperatures greater than the minimum $T_0$ for which this
solution exists. While the (gapped) SBH solution exists in the
interval $T_o\leq T \leq T_c$. At $T_c$ which corresponds to
$\tx={1\over 2}$, this solution goes over into the ungapped 
solution\footnote{Therefore the GWW transition identified in \cite{spenta}
with the black hole-string transition is the same as the
Douglas-Kazakov (DK) transition in our approach. A similar thing was
seen in \cite{gm} where the DK transition of $2d$ Yang-Mills was
mapped onto a GWW-like gapped to ungapped transition in terms of the
eigenvalues of Wilson loops.}.  The free energy in this phase is given
by,
\be F =- N^2 T \ltb \tx -
{1 \over 2} \ml[2 \tx] - {1\over 2}- b_1 \lb 1 - {1\over 4 \tx} \rb^4
\rtb.  \ee 
\ei

One of the points to note in our analysis is that the saddle
configurations for $u(h)$ are all of the same form as in the free
theory. It is only that $\vr$ is a different function of the
temperature. This is a mirror of the same phenomenon in the eigenvalue
density analysis that the functional form of the saddle configurations
of $\sigma(\theta)$ are not changed as one turns on the perturbative
coupling. This robust character of the saddle point configurations is
a positive indication in trying to extract universal features from
these results.

Finally, we should mention that the extension to non-zero charges also
follows in a straightforward way. As argued in \cite{basu}, the
effective matrix model has a logarithmic term which results in there
no longer being a uniform eigenvalue density saddle point
corresponding to thermal $AdS$.  In our approach, as argued in
Appendix \ref{appB} for a general matrix model of the form
Eq.\ref{weakacn}, we find the same saddle point equations as in the
eigenvalue analysis and thus the same phase diagram.


\section{Free Fermionic Phase Space Description}

\subsection{Relation between the \yt \ and Eigenvalue distributions} 
\label{secevyteqv}

Our analysis of the exact answer has been very different from the
usual eigenvalue analysis reviewed in Sec.\ref{secrev}.  It turns
out, rather remarkably, that there is nevertheless, a simple
relationship between the saddlepoint configurations $u(h)$, in both
the high and low temperature phases, with the corresponding
saddlepoint eigenvalue densities. We will now describe this relation.

Consider first the low temperature saddlepoint $u(h)=1$. The
corresponding saddlepoint for the eigenvalue density is
$\sigma(\theta)={1\over 2\pi}$. From the graph of these two
distributions, we notice that they are functional inverses of each
other (flip the horizontal and vertical axes of one to get the
other). In other words, we can make the identification
\ben\label{rel1} u &=& {\theta \over \pi} \nn 
{h \over 2 \pi} &=&
\sigma(\theta)\ .  
\een

This case may seem a little trivial, so let us consider the family of
saddlepoints that correspond to the unstable small black hole
Eq.\ref{usol2}
\be\label{usol22}
u(h)={1 \over \pi} \mt{\large cos}^{-1} \ltb {h-1 \over 2
\vr} \rtb {\hskip .8cm} h \in [\hm, \hma];  \ \ 2\vr\leq1\ .
\ee
together with $u(h)=1$ for $h\in [0,p]$. Applying relations \ref{rel1}
to this case, we get immediately $\sigma(\theta)={1\over
2\pi}(1+2\vr\cos\theta)$ which is the same as Eq.\ref{sigma2}.  Thus
the identification holds in this case. The two distributions are again
functional inverses of each other.\footnote{A similar relation was
also found in the case of $2d$ Yang-Mills theory on the
cylinder\cite{gm} though a phase space interpretation was not made.}

We finally come to the non-trivial saddlepoint corresponding to the
big black hole Eq.\ref{usol1}  
\be\label{usol12} u(h)= {1 \over \pi}
\mt{\large cos}^{-1} \ltb {h-1 \over 2\vr} + {\lb \vr - {1 \over 2}
\rb^2 \over 2 \vr \ h} \rtb \ \ \ \ h \in [\hm, \hma]\ .  
\ee

In this case, we have to be more careful. We see from the plot of
$u(h)$ that the functional inverse is ambiguous. For a given value of
$u$, there are two values of $h$.  This can be directly seen from the
fact that Eq.\ref{usol12} implies the quadratic relation
\be \label{bbhcurve}
h^2 - \ltb 1 + 2 \vr \cos\lb\pi u(h)\rb \rtb h + \lb \vr
-{1\over 2}\rb^2 =0\ . 
\ee
If we take the difference between the two solutions $h_+$ and $h_-$,
we obtain
\be\label{h+h-}
h_{+} - h_{-} = 2 \sqrt{2 \vr} \sqrt{1 - 2 \vr \mt{\large
sin}^2 \lb {\pi u \over 2} \rb} \mt{\large cos} \lb{\pi
u \over 2} \rb\ . 
\ee
We see that if we define 
\be
\mt{\large sin}^2 {\theta_0 \over 2} = {1 \over 2 \vr}
\ee
and modify the identifications Eq.\ref{rel1} to
\ben\label{rel2}
u &=& {\theta \over \pi} \nn
{h_+ -h_- \over 2 \pi} &=& \sigma(\theta)\ ,
\een
then we obtain precisely the eigenvalue distribution in Eq.\ref{sigma1}
\be
\sigma(\theta)={1 \over \pi \mt{\large sin}^2 \left ({\theta_0 \over
2} \right )} \sqrt{\mt{\large sin}^2 \left ({\theta_0 \over 2} \right
) - \mt{\large sin}^2\left ({\theta \over2}\right )} \mt{\large cos}
\left ({\theta \over 2} \right )\ .
\ee

\subsection{Fermionic Phase Space} \label{secfrefer}

The relations Eq.\ref{rel1} and \ref{rel2} between the saddlepoint
eigenvalue densities and the young tableau densities have a very
natural interpretation in terms of a free fermionic picture. This
fermionic picture is suggested by the fact that the eigenvalues of the
holonomy matrix behave like fermions. At the same time, the
representations of $U(N)$ also have an interpretation in the language
of non-interacting fermions with the number of boxes of the Young
tableaux being like the momentum (See \cite{douglas}, for
example). This suggests that the eigenvalue density is like a position
distribution while the Young Tableau density is like a momentum
distribution\footnote{In fact, $u(h)$ can be viewed as a plot of the
fermi distribution of momenta. The uniform density saddlepoint, for
instance, corresponds to a fully filled fermi sphere.}. Therefore, it is
natural to consider a phase space distribution which gives rise to
these individual distributions. In the classical (i.e. large $N$)
limit, we can describe this system of $N$ fermions in terms of an
incompressible fluid occupying a region of the two dimensional phase
space (see \cite{dhar} for a recent overview and references to the
large literature on the subject).

Therefore let us assume that the saddlepoints are all described by
some configuration in phase space, i.e.  some region $R$ of the two
dimensional plane such that the phase space density $\rho(h, \theta)$
obeys 
\ben\label{phspden} 
\rho(h,\theta) &=& {1\over 2\pi}\ ; \ \ \
(h,\theta)\in R \nn &=& 0 \ ; \ \ \ {\rm otherwise}.  
\een 
We can then
define the partial densities \footnote{The measure factor that appears
in Eq. \ref{defden} and Eq. \ref{mesurefac} suggests that $h$ is related to the usual polar coordinate
$r$ by $h={r^2\over 2}$. This redefinition is quite natural from the
point of view of free fermionic phase space where a similar change of
variables is made. See \cite{dhar} and \cite{grant}.}  
\ben\label{defden} 
u(h) &=&
\int_{-\pi}^{\pi} \rho(h,\theta)d\theta \nn \sigma(\theta) &=&
\int_0^{\infty} \rho(h,\theta)dh 
\een 
where the first integral is at
constant $h$ and the second at constant $\theta$.  Note that 
\be \label{mesurefac}
\int
\rho(h,\theta)dh d\theta =1\ .
\ee
\begin{figure}
\centering
\includegraphics[height=8cm]{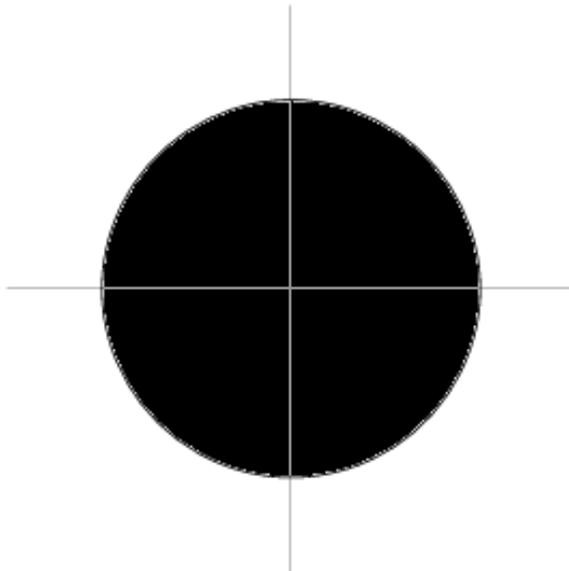}
\caption{Phase space distribution for Thermal AdS: $\vr$ =0 }
\label{thm_ads}
\end{figure}
\begin{figure}
\centering
\includegraphics[height=8cm]{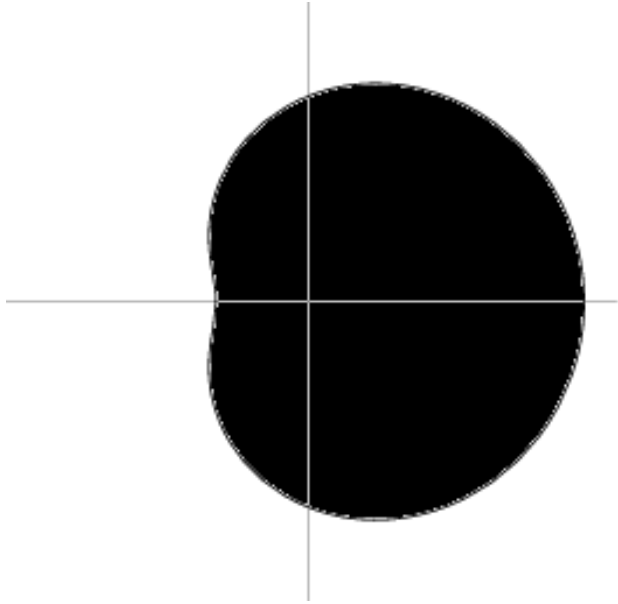}
\caption{Phase space distribution for SBH in ungapped phase: $\vr$ =0.4 }
\label{sbh}
\end{figure}
\begin{figure}
\centering
\includegraphics[height=9cm]{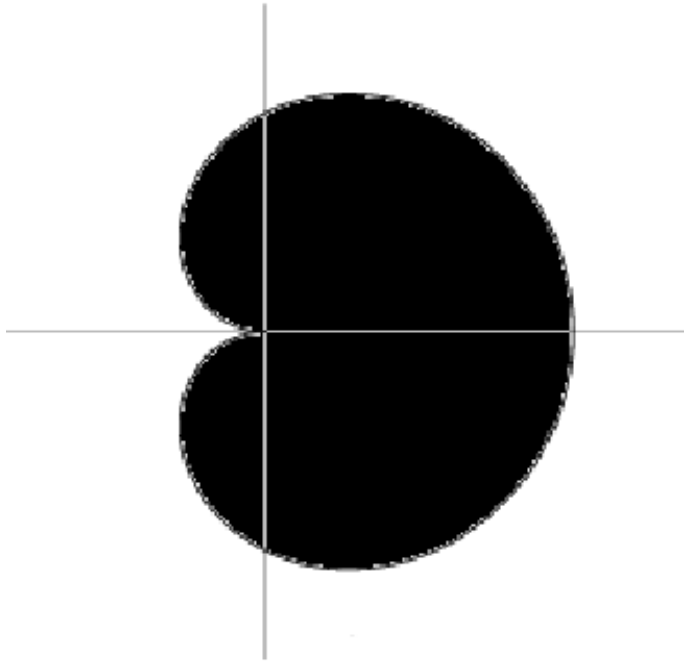}
\caption{Phase space distribution at GWW transition point : $\vr=0.5$ }
\label{gww}
\end{figure}
  
Let us now take the boundary of the region $R$ to be defined by a
curve\footnote{The symmetry of the effective action under $U\ra
U^{\dagger}$ implies the region is symmetric under $\theta \ra
-\theta$.}  $C(h,\theta)=0$.We now see that there can be different
situations depending on whether the solution $h=h(\theta)$ is single
valued or multiple valued. Thus for a single valued $h=h(\theta)$, we
see from Eqs.\ref{phspden} and \ref{defden} that the identification
Eq.\ref{rel1} follows. When we have multiple values for the solution
$h(\theta)$ then the identification between the different densities is
a little more non-trivial.  For instance, when we have two solutions
$h_{+}(\theta)$ and $h_{-}(\theta)$ with $h_+\geq h_-$, then it
follows from Eqs.\ref{phspden}, \ref{defden} that the relation between
the young tableaux density and the eigenvalue densities is that in
Eq.\ref{rel2}. Thus we have an interpretation for the relationships
that we observed in the previous subsection.

We therefore see that the large $N$ saddlepoints of the gauge theory
effective action, which correspond to the Thermal AdS, the small black
hole and the big black hole can each be thought of in terms of a
particular configuration in a free fermionic phase space.  There is a
particular shape associated to each of them. This shape, which is
determined by the curve $C(h,\theta)=0$, can, in general, only be
inferred from the knowledge of {\it both} $\sigma(\theta)$ and
$u(h)$. For the different saddlepoints that we have discussed, the
particular shapes are given in Figs.6-9.  We have plotted the
regions in polar coordinates $(r, \theta)$ after making the
identification in footnote 13.

\begin{figure}
\centering
\includegraphics[height=9cm]{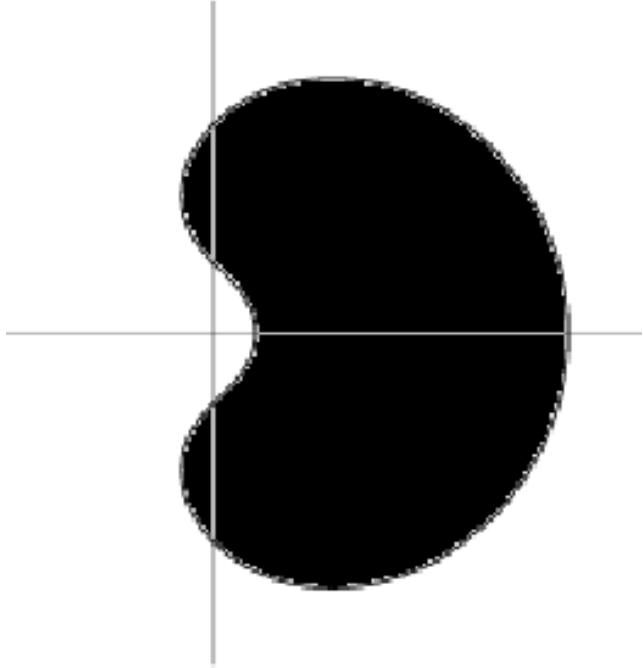}
\caption{Phase space distribution for SBH (in the gapped phase)/BBH}
\label{bbh}
\end{figure}

The shape corresponding to thermal $AdS$ is a disk $h(\theta) = 1$
(Fig. \ref{thm_ads}) which is exactly like that in the Lin, Lunin,
Maldacena description for global $AdS$.  For the big black hole we
have a shape Fig. \ref{bbh} which corresponds to a double valued
$h(\theta)$. The equation of the curve is given by, Eq. \ref{bbhcurve}
with $\pi u(h)$ replaced by $\theta$.  Note that the origin of phase
space is not contained in this region.  For the unstable saddlepoint
(SBH) we have the black hole regime where the shape is qualitatively
the same as the BBH, except that it is closer to the origin. At the
temperature $T_c$, the shape continuously changes to that in
Fig.\ref{gww}. The excited string state beyond this transition
occupies a region which includes the origin given by the curve
$h(\theta) = 1 + 2 \vr \cos\theta$ (Fig. \ref{sbh}).  We observe that
the shapes are qualitatively of two different kinds with the limiting
shape $h={r^2\over 2}=(1+\cos\theta)$ at the GWW transition point that
separates the two classes. Note that the boundary of the limiting
configuration is a separatrix between two different kinds of
trajectories in the fermi sea\footnote{We thank S. Wadia for useful
discussions on this point.}.

\section{Conclusions} \label{con}

We have studied thermal gauge theory by evaluating its partition
function at finite $N$. Taking the large $N$ limit of the full answer
gives us a new perspective on some already known facts about the phase
diagram of the gauge theory. The various phases and the transitions
between them can be viewed in terms of different dominant
representations of $U(N)$ that contribute to the partition
function. They are obtained as saddlepoints of an effective
functional, $S[u(h)]$ (like that given in Eq.\ref{Seffyt}) of a young
tableau density.  These representations can be viewed as excitations
of a one dimensional fermi sea whose ground state corresponds to
(thermal) AdS. In fact, a complete picture emerges when we combine the
saddlepoint representation $u(h)$ with the saddlepoint
$\sigma(\theta)$ of the eigenvalue density analysis. One obtains a
phase space picture which neatly combines both these answers into a
single two dimensional region with uniform density $\rho(h,\theta)$.
An important point to emphasise is that in general, one needs {\it
both} $u(h)$ and $\sigma(\theta)$ to obtain the curve $C(h,\theta)=0$
which delineates the region $R$ of phase space which is occupied. In a
way, as we have seen, $u(h)$ is more basic since it contains the
information to reconstruct $\sigma(\theta)$ through Eq.\ref{rel2} but
not vice versa.
  
This suggests that it is more natural to look for a description of the
dynamics directly in terms of the phase space density
$\rho(h,\theta)$. In particular, it would be nice to find a phase
space functional $S[\rho(h,\theta)]$ which reduces to the two
different functionals $S[u(h)]$ and to $S[\sigma(\theta)]$ on
appropriately integrating out. It would seem that the formalism
reviewed for example in \cite{dhar}, would be the appropriate one.

Such a formulation is likely to help in moving towards the goal of
reconstructing the local theory in the bulk with all its
redundancies. Note that the configurations we are considering in the
thermal history are all $S^3$ invariant. Thus the only non-trivial
directions in the bulk are those of the thermal $S^1$ and the radial
direction. It is tempting to identify these two directions with the
phase space directions of the fermions. In fact, the eigenvalues of
the fermions can be viewed as positions on the T-dual to the thermal
circle and the momenta should correspond to the radial direction as
per the usual AdS/CFT correspondence. As a first step it would be nice
to see how the topology of the bulk is encoded in the geometry of the
phase space regions corresponding to the different saddlepoints. Note
that we argued that the shape of the regions is fairly robust against
coupling effects and so should be telling us something generic about
the bulk geometries even when they are subject to all kinds of
$\alpha^{'}$ corrections.

It may seem a little confusing to talk about phase space without
talking about time. But since we are describing euclidean gravity 
configurations in terms of euclidean gauge theory, time does not 
enter directly. A closely similar situation arises in the study of 
Euclidean $2d$ Yang-Mills theory on the torus. Its free fermion 
representation was interpreted \cite{baby} in terms of contributions from 
different "baby universe" saddle points of the Euclidean gravity 
partition function of the dual geometry (in this case a 4d extremal black 
hole in an ${\cal N}=2$ theory)\cite{vafa}. 

As mentioned in the introduction, the LLM scenario was one of the
inspirations for many of these ideas. It would be nice to see if there
was a concrete connection to be made between those cases and those
studied here. At first sight they seem to be very different. However
one possible link between them is that unitary matrix models very
similar to those studied here give the partition functions that count
BPS states. In fact, it would be nice to apply our finite $N$ answers
for these models to these countings\footnote{We thank S. Minwalla for
this suggestion.}.

Another interesting direction to generalise might be to the gauge
theory on spatial sections which are for instance, $S^1\times R^2$
rather than $S^3$. We now have a matrix quantum mechanics
\cite{wiseman}.  Toy matrix models for this case have been studied for
example in \cite{bobby}.  In these cases a free fermionic description
has been employed to study the matrix quantum mechanics.



{\vskip 2cm}

\noindent
{\bf Acknowledgement:} We benefitted very much from conversations with
J. R. David, A. Dhar, S. Minwalla, K. S. Narain and S. Wadia. We also
appreciate helpful comments by O. Aharony, S. Minwalla and S. Wadia on  
the draft. One of
us (R.G.) also acknowledges the hospitality of IFT, Sao Paulo, where a
portion of this work was completed. We also thank the participants of the
ISM07 workshop on string theory for their comments. 
Finally, we are both beholden to the people
of India for the unstinting support lent to fundamental research.
\bc
{\bf --------------------------------------------------------}
\ec
\newpage

\appendix
\noindent
{\Large {\bf Appendix}}

\section{Details of the Evaluation of the Free 
Energy} \label{appa}

\subsection{Free Energy for solution class 1}

The effective action is given by 
\be
-\seff = \int_{h_L}^{h_U} dh {-\hskip -12pt}\int_{h_L}^{h_U} dh'
u(h) u(h') \ml |h - h'| - 2\int_{h_L}^{h_U} dh \
u(h) \ h \ml(h) 
+k'  +  1 + k'  \ml (a_1k')
\ee
For solution class 1, $h_L=q$ and $h_U=p$. $u(h)$ has support only for
$q\leq h\leq p$.

Using the following saddle equation of motion we can replace the double
integration by a single integration,
\be\label{c1}
{-\hskip -12pt}\int_q^p dh' \ln[h-h']\ u(h') = h \ln \ltb{h \over \xi}
\rtb - h + {\cal C}_1\ .
\ee
Hence the effective action becomes,
\be
-\seff^0 = - \int_q^p dh h u(h) \ln[h] + \vr \ln[\vr] - {3 \over 4} \ln
 [\vr] + {1\over2} + {\cal C}_1\ .
\ee
The constant ${\cal C}_1$ can be evaluated from the equation \ref{c1}
at $h= p$.
\be
{\cal C}_1 = \hma + \hma \ln \ltb {\hma \over \vr} \rtb + \int_q^p dh'
\ln[\hma-h']\ u(h')\ .
\ee
After some algebra we find,
\be
\int_q^p dh h u(h) \ln[h] = \vr \ln[\vr] + {1\over4} \ln[\vr] +
{1\over 2}
\ee
and 
\be
{\cal C}_1 = {1\over2} \ln [\vr/2] + \vr - {1\over2}\ .
\ee
Finally we get,
\ben
\seff^0 &=& \ln[\vr] - {\cal C}_1 \nn
&=& - \ltb \vr - {1\over2}\ln[2\vr] - {1\over2}\rtb \ .
\een

\subsection{Free Energy for solution class 2}
The effective action for the solutions in class 2 is given by,
\ben 
-\seff = \int_{0}^{\hma} dh {-\hskip
  -12pt\int}_{0}^{\hma} &dh'& u(h)u(h') \ml |h -
h'| - 2\int_{0}^{\hma} dh u(h) h\ml(h) \nn
&+& k' + 1 + k' \ml (a_1 k').
\een
Breaking the $h$ integration into two pieces, we can write 
the effective action as,
\ben \label{seffb2}
-\seff &=& \int_{0}^{\hm} dh{-\hskip
  -12pt\int}_{0}^{\hma} dh' u(h) u(h')\ml |h -
h'| 
+ \int_{\hm}^{\hma} dh {-\hskip
  -12pt\int}_{0}^{\hm} dh' u(h) u(h') \ml |h -
h'| \nn
&-&  2\int_{0}^{\hma} dh u(h)h\ml(h)
+ k'+ 1 + k'\ml (a_1k').
\een
In the second term on the right hand side, one can use the saddle equation
Eq. (\ref{sadeq2}) with $h_L=0$ and $h_U = \hma$. This gives
\be
\int_{\hm}^{\hma} dh {-\hskip -12pt\int}_{0}^{\hma} dh' u(h) 
u(h') \ml |h - h'| = \int_{\hm}^{\hma} dh u(h) \lb
h\ml[h] - h\ml[\vr] -h-{\cal C}_2 \rb,
\ee
where the constant ${\cal C}_2$ is given by,
\be
{\cal C}_2 = \hma - \hma \ \ml\ltb{\hma \over \vr} \rtb + {-\hskip
-12pt \int}_{0}^{\hma} dh' \ u(h') \ \ml|h-h'| \ .
\ee
After some algebra we get,
\be
{\cal C}_2 = \ml [\vr]\ .
\ee
Therefore
\ben
\int_{\hm}^{\hma} dh {-\hskip -12pt\int}_{0}^{\hma} dh' u(h) 
u(h') \ml |h - h'| &=& \int_{\hm}^{\hma} dh h 
u(h) \ml[h] \nn 
&-& (1 + \ml[\vr]) \lb k' + {1 \over 2} - {\hm^2
\over 2} \rb + {\cal C}_2(1-\hm).
\een

Calculating the first term on the right hand side of
Eq. (\ref{seffb2}) we get, 
\ben 
\int_{0}^{\hm} dh {-\hskip
-12pt\int}_{0}^{\hma} dh' u(h) u(h') \ml |h - h'| &=&
-{\hm^2 \over 2} + \hm^2 \ml[\hm] + \int_{\hm}^{\hma} dh' h' u(h')
\ml[h'] \nn &-& \hm + \int_{\hm}^{\hma} dh'u(h') (\hm -h')
\ml[h'-\hm]\ .  
\een 
Calculating other terms in on the right hand side
of Eq. (\ref{seffb2}) we finally see that the on-shell effective
action in second branch vanishes. Hence the free energy in this branch
is zero.

\section{A Class of General Matrix Model Actions}\label{appB}

In this appendix we will generalize the result of section \ref{seclargeN}
to a generic effective action which is a function of
$\mtr U \mtr U^{\dagger}$. We will expand the effective action in a
power series in $x$, where $x=\mtr U \mtr U^{\dagger}$.
\be 
-\seff = \sum \limits_{n=1} {a_n x^n \over N^{2(n-1)}} = N^2 \sum
\limits_{n=1} {a_n x^n \over N^{2n}}.
\ee
Using this form of $\seff$, we can write
\be
\me^{-\seff} = \prod \limits_i^m \sum \limits_{k_i=0}^{\infty} 
{(a_i)^{k_i}\ x^{(\sum_{i=1}^{\infty}i k_i)} \over k_i! 
N^{2(i-1)}}.
\ee 
The \pf \ is given by,
\be
Z= \int [dU] \me^{-\seff},
\ee
can be evaluated using the methods of Sec.3 and Sec.4,
\be
Z = \sum \limits_{\{k_i\}=0}^{\infty} {a_1^{k_1} 
\cdot \cdot a_m^{k_m}
\over k_1! \cdot \cdot k_m!} \ltb {\sum \limits_R d_R^2(S_k) 
\over N^{2
\sum \limits_{l=1}^{m} k_l(l-1)}}\rtb \ \ ,\ \  k= \sum
\limits_{l=1} l k_l \ .
\ee
Define,
\ben
f(k) &=& \sum \limits_{\{k_i\}=0}^{\infty} {a_1^{k_1} 
\cdot \cdot a_m^{k_m} \cdot
\over k_1! \cdot \cdot k_m!\cdot} {k! \over N^{2
\sum \limits_{l=1} k_l(l-1)}} \nn
&=& \sum \limits_{\{k_i\}=0}^{\infty} \me^{G(\{k_i\})} \ ,
\een
where
$G(\{k_i\})$ is given by,
\be
G(\{k_i\}) = \sum \limits_{i=1} k_i \ml[a_i] - \sum
\limits_{i=1}^{m} \lb k_i\ml [k_i] - k_i\rb + k\ml [k] -k - 2 
\sum \limits_{i=1}^{m}k_i(i-1) \ml[N].
\ee
In the large $N$ limit the \pf receives its dominant 
contribution only from
the extremum value of $G$. Minimising $G$ with respect 
to $k_i$, remembering to introduce a Lagrange multiplier for 
the constraint $\sum
\limits_{i=1}ik_i = k$, we
get,
\be
k_i^{\prime} = a_i\beta^i \ ,
\ee
where $k_i^{\prime} = {k_i \over N^2}$ and $\beta = 
\me^{-(\alpha + 2
\ln N)}$ where $\alpha$ is the Lagrange 
multiplier enforcing the
constraint. 

$\beta$ is determined through the following relation,
\ben
k^{\prime} &=& \sum \limits_{i=1}^{\infty}  i k_i' \nn
&=& \sum \limits_{i=1}i a_i \beta^i \nn
&=& \beta \tilde S^{\prime} (\beta) \ ,
\een
where $\tilde S^{\prime} (\beta)$ is given by,
\be
\tilde S(\beta) = \sum \limits_{i=1} a_i \beta^i\ .
\ee
Once we fix the undetermined multiplier, then $G$ 
can be written as,
\be
{G \over N^2} = k'\ml [k'] - k' - k' \ml[\beta] + 
\tilde S(\beta)\ .
\ee
Hence $f(k)$ is given (upto multiplicative factors 
which are unimportant in the large $N$ limit).
\be
f(k)\propto \me^{N^2 (k'\ml[{k'\over \beta}]-1)} 
\me^{N^2 \tilde S(\beta)}\ .
\ee
So the \pf can be written as,
\be
Z = \sum \limits_{k=0}^{\infty} {\tilde a^k \over k!}
\sum \limits_R d_R^2(S_k) \ ,
\ee
where $\tilde a^k= \me^{k \ln[k'/\beta] - k 
+ N^2 \tilde S(\beta)}$. Like before we will write the 
\pf in the following form,
\be
Z= \int [dh(x)] \exp[- N^2 \seff],
\ee
where $\seff$ is given by,
\ben
- \seff &=& {-\hskip -12pt}\int_0^1 dx \int_0^1 dy \ln
|h(x)-h(y)| \nn &-& 2\int_0^1 dx h(x) \ln h(x)
\ +\ k' \ln\lb{\tilde a\ k'}\rb +k'+1\ . 
\een
Hence the saddlepoint equation is given by,
\be
{-\hskip -12pt}\int_{h_L}^{h_U} dh' {u(h') \over h  -  h'} 
= \ln({h\over \tilde \vr}),
\ee
where $\tilde \vr$ is given by,
\be
\tilde \vr^2 = k' \tilde S'(\beta)\ .
\ee 
For this generic effective action the solutions of 
the saddle equation are given by,
\ben
\beta\tilde S'(\beta) &=& \beta {\hskip 2.3cm}  
\mt{\large for} \ \ \beta \in [0,{1 \over 4}] \nn
\tilde S'(\beta) &=& {1 \over 4 \sqrt{\beta}
(1-\sqrt{\beta})} \ \ \mt{\large for}\ \ 
\beta \in [{1 \over 4} ,1] \ .
\een
For $\beta \in [0,{1 \over 4}]$, $\xi 
= \sqrt{\beta}$ and for $\beta \in 
[{1 \over 4} ,1]$, $\xi = {1 \over 4 (1- \sqrt{\beta})}$.

These equations are identical to those obtained by the Hartree-Fock
analysis of eigenvalue density Eq. \ref{revsol1} and Eq. \ref{revsol2}.


\begin{thebibliography}{99}

\bibitem{llm} H.~Lin, O.~Lunin and J.~M.~Maldacena, ``Bubbling AdS
  space and 1/2 BPS geometries,'' JHEP {\bf 0410}, 025 (2004)
  [arXiv:hep-th/0409174].  

\bibitem{mandal}
  G.~Mandal,
  ``Fermions from half-BPS supergravity,''
  JHEP {\bf 0508}, 052 (2005)
  [arXiv:hep-th/0502104].



\bibitem{grant}
  L.~Grant, L.~Maoz, J.~Marsano, K.~Papadodimas and V.~S.~Rychkov,
  ``Minisuperspace quantization of 'bubbling AdS' and free fermion  droplets,''
  JHEP {\bf 0508}, 025 (2005)
  [arXiv:hep-th/0505079].


\bibitem{witten}
  E.~Witten,
  ``Anti-de Sitter space, thermal phase transition, and confinement in  gauge
  theories,''
  Adv.\ Theor.\ Math.\ Phys.\  {\bf 2}, 505 (1998)
  [arXiv:hep-th/9803131].

\bibitem{hp}
  S.~W.~Hawking and D.~N.~Page,
  ``Thermodynamics Of Black Holes In Anti-De Sitter Space,''
  Commun.\ Math.\ Phys.\  {\bf 87}, 577 (1983).

\bibitem{sundborg}
  B.~Sundborg,
  ``The Hagedorn transition, deconfinement and N = 4 SYM theory,''
  Nucl.\ Phys.\  B {\bf 573}, 349 (2000)
  [arXiv:hep-th/9908001].

\bibitem{hallin}
  J.~Hallin and D.~Persson,
  ``Thermal phase transition in weakly interacting, large N(c) {QCD},''
  Phys.\ Lett.\  B {\bf 429}, 232 (1998)
  [arXiv:hep-ph/9803234].

\bibitem{shiraz}
  O.~Aharony, J.~Marsano, S.~Minwalla, K.~Papadodimas and M.~Van Raamsdonk,
  ``The Hagedorn / deconfinement phase transition in weakly coupled large N
  gauge theories,''
  Adv.\ Theor.\ Math.\ Phys.\  {\bf 8}, 603 (2004)
  [arXiv:hep-th/0310285].

\bibitem{hong}
  H.~Liu,
  ``Fine structure of Hagedorn transitions,''
  arXiv:hep-th/0408001.

 \bibitem{spenta}
  L.~Alvarez-Gaume, C.~Gomez, H.~Liu and S.~Wadia,
  ``Finite temperature effective action, AdS(5) black holes, and 1/N
  expansion,''
  Phys.\ Rev.\  D {\bf 71}, 124023 (2005)
  [arXiv:hep-th/0502227].

\bibitem{azuma}
  Takehiro Azuma, Pallab Basu, Spenta R. Wadia,
``Monte Carlo Studies of the GWW Phase Transition in Large-N Gauge
 Theories,'' arXiv:0710.5873[hep-th].

\bibitem{dk}
  M.~R.~Douglas and V.~A.~Kazakov,
  ``Large N phase transition in continuum QCD in two-dimensions,''
  Phys.\ Lett.\  B {\bf 319}, 219 (1993)
  [arXiv:hep-th/9305047].


\bibitem{ksw}
  V.~A.~Kazakov, M.~Staudacher and T.~Wynter,
  ``Character expansion methods for matrix models of dually weighted graphs,''
  Commun.\ Math.\ Phys.\  {\bf 177}, 451 (1996)
  [arXiv:hep-th/9502132].

\bibitem{gm}
  D.~J.~Gross and A.~Matytsin,
  ``Some Properties Of Large N Two-Dimensional Yang-Mills Theory,''
  Nucl.\ Phys.\  B {\bf 437}, 541 (1995)
  [arXiv:hep-th/9410054]. 

\bibitem{gw}
  D.~J.~Gross and E.~Witten,
  ``Possible Third Order Phase Transition In The Large N Lattice Gauge
  Theory,''
  Phys.\ Rev.\  D {\bf 21} (1980) 446.

\bibitem{wadia}
  S.~R.~Wadia,
  ``N = Infinity Phase Transition In A Class Of Exactly Soluble Model Lattice
  Gauge Theories,''
  Phys.\ Lett.\  B {\bf 93}, 403 (1980).

 
\bibitem{spenta2}
  L.~Alvarez-Gaume, P.~Basu, M.~Marino and S.~R.~Wadia,
  ``Blackhole / string transition for the small Schwarzschild blackhole of
  AdS(5) x S**5 and critical unitary matrix models,''
  Eur.\ Phys.\ J.\  C {\bf 48}, 647 (2006)
  [arXiv:hep-th/0605041].





\bibitem{hamermesh}
Morton Hamermesh,
``Group Theory and its Application to Physical Problems,'' 
  Dover publications.

\bibitem{lasalle}
Michel Lasalle,
"`Explicitation of Characters of the Symmetric Group,"'
C. R. Acad. Sci. Paris, Ser I{\bf 341}, 529-534 (2005).



\bibitem{marino}
  M.~Marino,
  ``Les Houches lectures on matrix models and topological strings,''
  arXiv:hep-th/0410165.





\bibitem{basu}
  P.~Basu and S.~R.~Wadia,
  ``R-charged AdS(5) black holes and large N unitary matrix models,
  Phys.\ Rev.\  D {\bf 73}, 045022 (2006)''
  [arXiv:hep-th/0506203].

\bibitem{yamada}
  D.~Yamada and L.~G.~Yaffe,
  ``Phase diagram of N = 4 super-Yang-Mills theory with R-symmetry chemical
  potentials,''
  JHEP {\bf 0609}, 027 (2006)
  [arXiv:hep-th/0602074].

  T.~Harmark and M.~Orselli,
  ``Quantum mechanical sectors in thermal N = 4 super Yang-Mills on R x
  S**3,''
  Nucl.\ Phys.\  B {\bf 757}, 117 (2006)
  [arXiv:hep-th/0605234].

\bibitem{harmark}
  T.~Harmark, K.~R.~Kristjansson and M.~Orselli,
  JHEP {\bf 0709}, 115 (2007)
  [arXiv:0707.1621 [hep-th]].

\bibitem{douglas}
  M.~R.~Douglas,
  ``Conformal field theory techniques in large N Yang-Mills theory,''
  arXiv:hep-th/9311130.



\bibitem{arsiwalla}
  X.~Arsiwalla, R.~Boels, M.~Marino and A.~Sinkovics,
  ``Phase transitions in q-deformed 2d Yang-Mills theory and topological
  strings,''
  Phys.\ Rev.\  D {\bf 73}, 026005 (2006)
  [arXiv:hep-th/0509002].

\bibitem{baby}
  R.~Dijkgraaf, R.~Gopakumar, H.~Ooguri and C.~Vafa,
  ``Baby universes in string theory,''
  Phys.\ Rev.\  D {\bf 73}, 066002 (2006)
  [arXiv:hep-th/0504221].

\bibitem{vafa}
  C.~Vafa,
  ``Two dimensional Yang-Mills, black holes and topological strings,''
  arXiv:hep-th/0406058.

\bibitem{wiseman}
  O.~Aharony, J.~Marsano, S.~Minwalla, K.~Papadodimas, M.~Van Raamsdonk
  and T.~Wiseman,
  ``The phase structure of low dimensional large N gauge theories on tori,''
  JHEP {\bf 0601}, 140 (2006)
  [arXiv:hep-th/0508077].

\bibitem{bobby}
  P.~Basu, B.~Ezhuthachan and S.~R.~Wadia,
  ``Plasma balls / kinks as solitons of large N confining gauge theories,''
  JHEP {\bf 0701}, 003 (2007)
  [arXiv:hep-th/0610257].









\bibitem{dhar}
  A.~Dhar,
  ``Bosonization of non-relativstic fermions in 2-dimensions and collective
  field theory,''
  JHEP {\bf 0507}, 064 (2005)
  [arXiv:hep-th/0505084].



\bibitem{raju}
  J.~Kinney, J.~M.~Maldacena, S.~Minwalla and S.~Raju,
  ``An index for 4 dimensional super conformal theories,''
  Commun.\ Math.\ Phys.\  {\bf 275}, 209 (2007)
  [arXiv:hep-th/0510251].






  

\end{thebibliography}
\end{document}